\documentclass[american,aps,manuscript,aps,preprint]{revtex4-1}
\usepackage[T1]{fontenc}
\usepackage[latin9]{inputenc}
\usepackage{array}
\usepackage{multirow}
\usepackage{amsmath}
\usepackage{graphicx}

\makeatletter

\providecommand{\tabularnewline}{\\}

 \@ifundefined{textcolor}{}
 {%
   \definecolor{BLACK}{gray}{0}
   \definecolor{WHITE}{gray}{1}
   \definecolor{RED}{rgb}{1,0,0}
   \definecolor{GREEN}{rgb}{0,1,0}
   \definecolor{BLUE}{rgb}{0,0,1}
   \definecolor{CYAN}{cmyk}{1,0,0,0}
   \definecolor{MAGENTA}{cmyk}{0,1,0,0}
   \definecolor{YELLOW}{cmyk}{0,0,1,0}
 }

\makeatother

\usepackage{babel}
\begin{document}

\title{Resonant production of spin-3/2 color octet electron at the LHeC}

\author{M. \c{S}ahin}

\email{mehmet.sahin@usak.edu.tr}

\address{Usak University, Department of Physics, U\c{s}ak, Turkey.}
\begin{abstract}
In this work, we investigate resonant production of spin-3/2 color
octet electron at the Large Hadron electron Collider (LHeC). Signal
and background analysis are performed and discovery, observation and
exclusion limits are determined for spin 3/2 color octet electron
masses. Reachable values of the compositeness scale are presented
as a function of the spin-3/2 color octet electron masses.
\end{abstract}
\maketitle

\section{INTRODUCTION}

Standard Model (SM) of particle physics successfully explains electromagnetic,
weak and strong interactions of fundamental particles. SM predictions
are consistent with numerous experiments. Furthermore, results of
the ATLAS and the CMS Collaborations \cite{ATLAS,CMS} are consistent
with the SM Higgs boson hypothesis. If the new particle is the SM
Higgs boson, why the fundamental particles have mass is explained
by SM. However, SM does not explain some cases such as quark-lepton
symmetry, family replication, hierarchy problems, charge quantization,
etc. Many theoretical hypothesis are proposed to clarify these cases.
Supersymmetry (SUSY), extra dimensions, grand unified theories (GUTs)
and compositeness seem to be most promising candidates for beyond
the SM physics. 

The subjects of family replication and quark-lepton symmetry are explained
in the best manner by compositeness. In the framework of composite
models quarks, leptons and gauge bosons are composite particles made
up of more basic constituents. These basic constituents are named
preons. The preonic models lead to a rich spectrum of new type particles
such as diquarks, leptoquarks, leptogluons, dileptons and excited
fermions etc. 

We have interested in color octet leptons which are predicted in all
composite models with colored preons \cite{souza,Harari,Fritzch,Greenberg,Barbieri,Baur,Celikel}.
For example, in the framework of fermion-scalar models, leptons would
be a bound state of one fermionic preon and one scalar antipreon $l=(F\bar{S})=1\oplus8$
(both F and S are color triplets), then each SM lepton is thought
to accompany with its own color octet partner \cite{Celikel}. Color
octet electrons have the same status as the excited leptons. As the
excited leptons, color octet electrons have a spin of 1/2 or 3/2.
The motivation for spin-3/2 leptons comes from composite models \cite{Lopes,Tosa}
and supergravity gauge theories \cite{Nieuw}. Spin-1/2 color octet
electron are investigated in earlier papers \cite{Rizzo1,Rizzo2,Streng,Celikel,Hewett,Kantar}.
Recently spin-1/2 color octet electron have been analyzed for future
high energy colliders: Large Hadron electron Collider (LHeC) \cite{Sahin},
International Linear Collider (ILC) and Compact Linear Collider (CLIC)
\cite{Akay} and Large Hadron Collider (LHC) \cite{Mandal}. Although
spin-3/2 color octet electrons are mentioned in literature, they have
not been investigated in details up to now. In order to close this
gap in the literature, we investigate resonant production of the spin-3/2
color octet electron at the LHeC. 

LHeC project supported by CERN, ECFA and NuPECC is a new LHC based
electron-proton and electron-ion collider \cite{LHeCweb}. LHeC is
under design for synchronous operation with the LHC at CERN in the
twenties. Conceptual design report (CDR) of the LHeC is completed
and published last summer \cite{LHeCCDR}. The CDR of the LHeC describes
two options for the LHeC, a ring-ring (RR) and a linac-ring (LR) configuration.
After the CDR was completed, it was decided to continue the technical
design work for the linac-ring (LR) configuration in a recent workshop
\cite{ECFA,LHeCLR}. History and status of linac-ring type collider
proposals can be found in review \cite{Sultansoy}. We have preferred
two energy options of linac-ring (LR) type LHeC in our calculations.
The first one is lower-energy LHeC ERL option (see LHeC CDR \cite{LHeCCDR}
section 7.1.2). The lepton beam energy is $60$ GeV and proton beam
energy is $7000$ GeV in this option. Electron-proton luminosity is
$10{}^{33}\, cm^{-2}\, s^{-1}$. The latter option is higher-energy
LHeC ERL option (see LHeC CDR \cite{LHeCCDR} section 7.1.5). The
lepton beam energy is 150 GeV and proton beam energy is $7000$ GeV
in this option. Electron-proton luminosity is $10{}^{35}\, cm^{-2}\, s^{-1}$.
These properties of LHeC are shown in Table 1.

\begin{table}
\begin{tabular}{|c|c|c|c|}
\hline 
Stage & $E{}_{e}$, GeV & $\sqrt{s}$, TeV & L, $10{}^{33}\, cm^{-2}\, s^{-1}$ \tabularnewline
\hline 
\hline 
Lower-energy LHeC ERL (LHeC-1) & $60$ & $1.296$ & $1$\tabularnewline
\hline 
Higher-energy LHeC ERL (LHeC-2) & $150$ & $2.049$ & $100$\tabularnewline
\hline 
\end{tabular}

\caption{Tentative parameters of the LHeC linac-ring options. ERL denotes energy
recovery linac.}
\end{table}

In this paper, we consider resonant production of spin-3/2 color octet
electron ($e_{8}$) at the Large Hadron electron Collider. In the
Section II, the interaction Lagrangian of spin-3/2 leptogluons, decay
widths and resonant production cross sections of spin-3/2 color octet
electron at different stages of the LHeC are presented. In Section
III, signal and background analysis are performed at the LHeC and
achievable masses and compositeness scale are determined for spin-3/2
color octet electrons. Finally, we give concluding remarks in Section
IV.

\section{INTERACTION LAGRANGIAN, DECAY WIDTH AND CROSS SECTIONS}

For the interaction of spin-3/2 leptogluons with corresponding lepton
and gluon we use following Lagrangian:

\begin{equation}
L=\frac{1}{\Lambda^{2}}\underset{l}{\sum}\left\{ \bar{l}_{8}^{\alpha}\, g_{s}\,\partial^{\mu}\, G_{\mu\nu}^{\alpha}\sigma^{\mu\nu}(\eta_{L}l_{L}+\eta_{R}l_{R})+h.c.\right\} ,
\end{equation}

where $\bar{l}_{8}^{\alpha}$ is a Rarita-Schwinger vector spinor
\cite{Rarita} field for spin-3/2 leptogluons, index $\alpha=1,\,2,...,8$
denotes color, $g_{s}$ is gauge coupling, $G_{\mu\nu}^{\alpha}$
is the field strength tensor for gluon, $\sigma{}^{\mu\nu}$ is the
anti-symmetric tensor, $\eta{}_{L}$ and $\eta_{R}$ are the chirality
factors, $l{}_{L}$ and $l_{R}$ represent left and right spinor components
of leptons and $\Lambda$ is the compositeness scale. We introduce
only the first family leptogluons in this articles. We assume that
the leptogluon of the first family will be lighter than the others
following what we observe experimentally in SM. The leptonic chirality
conservation requires $\eta_{L}\eta_{R}=0$ and following the almost
exclusively left-handed neutrinos, that we observe in the nature,
we set $\eta_{L}=1$ and $\eta_{R}=0$ in our calculations. Propagator
of spin-3/2 color octet electron is represented by equation 2.

\begin{equation}
P^{\mu\nu}=\frac{1}{p^{2}-M_{e_{8}}^{2}}[-(\gamma^{\mu}p_{\mu}+M_{e_{8}})(g^{\mu\nu}-\frac{p^{\mu}p^{\nu}}{M_{e_{8}}^{2}})-\frac{1}{3}(\gamma^{\mu}+\frac{p^{\mu}}{M_{e_{8}}})(\gamma^{\mu}p_{\mu}-M_{e_{8}})(\gamma^{\mu}+\frac{p^{\nu}}{M_{e_{8}}})]
\end{equation}

Decay width of spin-3/2 color octet electrons is given by 

\begin{equation}
\Gamma_{e_{8}}=\frac{\alpha_{s}M_{e_{8}}^{5}}{12\Lambda^{4}}.
\end{equation}

The analytic expression for partonic level differential cross-section
for the process $e^{-}p\rightarrow e_{8}\rightarrow ge^{-}$ is given
by

\begin{equation}
\frac{d\hat{\sigma}}{d\hat{t}}(e^{-}p\rightarrow e_{8}\rightarrow ge^{-})=-\frac{64\alpha_{s}^{2}\hat{s}\hat{t}(\frac{\hat{s}^{2}}{M_{e_{8}}^{2}}-3(\hat{s}+\hat{t}))^{2}\pi}{\alpha_{s}^{2}M_{e_{8}}^{12}+144\Lambda^{8}(M_{e_{8}}^{2}-\hat{s})^{2}}.
\end{equation}

where $\hat{s}$ and $\hat{t}$ are Mandelsam variables, $\alpha_{s}$
is strong coupling constant and $\hat{\sigma}$ is partonic cross-section. 

For the numerical calculations we implement this Lagrangian into the
CALCHEP program \cite{Pukhov,Belyaev}. Figure 1 presents decay widths
of spin-3/2 color octet electrons for $\Lambda=M_{e_{8}}$ , $\Lambda=5$
TeV and $\Lambda=10$ TeV.

\begin{figure}
\includegraphics[scale=0.7]{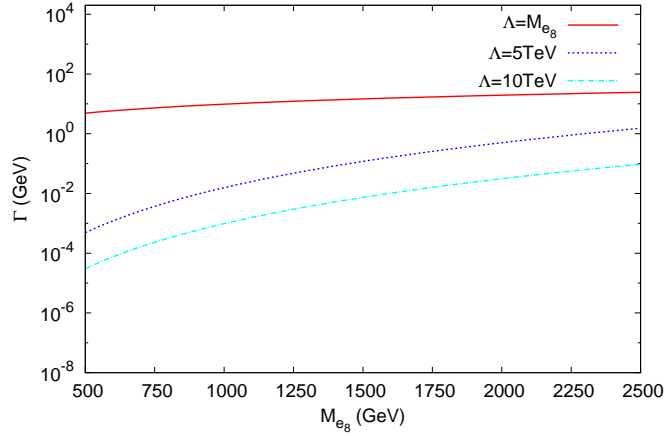} 

\caption{Spin-3/2 color octet electron decay width via its mass for $\Lambda=M_{e_{8}}$,
$\Lambda=5$ TeV and $\Lambda=10$ TeV.}
\end{figure}

Resonant production cross sections of spin-3/2 color octet electrons
for the two stages of the LHeC which are in Table 1, are calculated
using CALCHEP with CTEQ6L \cite{CTEQ6L} parton distribution functions.
In these calculations, we used factorization scale $Q^{2}=M_{e_{8}}^{2}$.
Figure 2 presents resonant production cross sections of spin-3/2 color
octet electrons for $\Lambda=M_{e_{8}}$, $\Lambda=5$ TeV and $\Lambda=10$
TeV at the LHeC-1 with $\sqrt{s}=1.296$ TeV. It is seen from Figure
2 that spin-3/2 color octet electrons have sufficiently high cross-section
values for $\Lambda=M_{e_{8}}$ and $\Lambda=5$ TeV even at the high
mass values of $e_{8}$. Figure 3 shows resonant production cross
section of spin-3/2 color octet electrons for $\Lambda=M_{e_{8}}$,
$\Lambda=5$ TeV and $\Lambda=10$ TeV at the LHeC-2 with $\sqrt{s}=2.049$
TeV. It is seen from Figure 3 that spin-3/2 color octet electrons
have sufficiently high cross-section for $\Lambda=M_{e_{8}}$, $\Lambda=5$
TeV, and $\Lambda=10$ TeV at the high mass values of $e_{8}$. 

\begin{figure}
\includegraphics[scale=0.7]{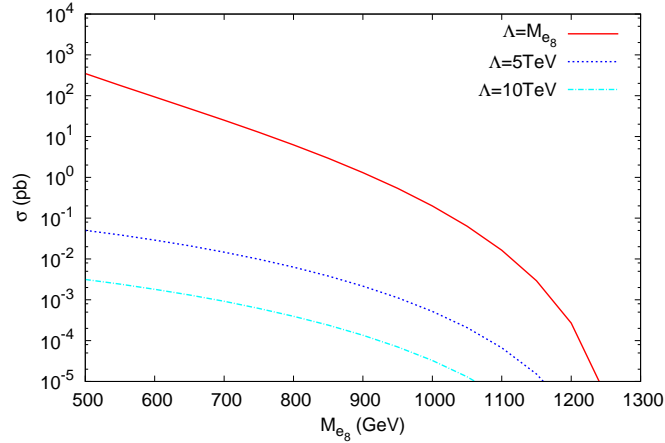}

\caption{Resonant production of spin-3/2 color octet electrons at the LHeC-1
with $\sqrt{s}=1.296$ TeV.}
\end{figure}

\begin{figure}
\includegraphics[scale=0.7]{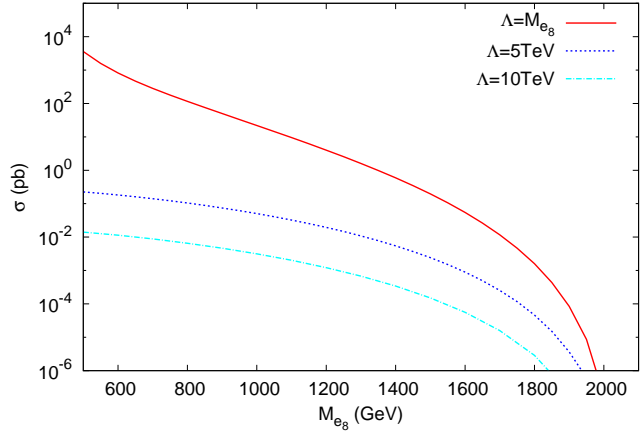}

\caption{Resonant production of spin-3/2 color octet electrons at the LHeC-2
with $\sqrt{s}=2.049$ TeV.}
\end{figure}

\section{SIGNAL AND BACKGROUND ANALYSIS}

\subsection{Lower-energy LHeC ERL stage (LHeC-1 with $\sqrt{s}=1.296$ TeV and
$L_{int}=10\, fb^{-1}$)}

The center of mass energy of this LHeC option is $1.296$ TeV and
integrated luminosity is $L_{int}=10\, fb^{-1}$. Our signal process
is $ep\rightarrow e_{8}\rightarrow eg+X$ and background process is
$ep\rightarrow ej+X$ through $\gamma$ and $Z$ exchange, where g
denotes gluon and $j$ represents jets which are composed of quarks
($u,$ $\bar{u}$, $d,$ $\bar{d}$, $c$, $\bar{c}$, $s$, $\bar{s}$,
$b$, $\bar{b}$ ) for background. In order to determine appropriate
cuts we need transverse momentum ($P_{T}$ ) and pseudo-rapidity ($\eta$)
distributions of signal and background processes.

Figure 4 presents normalized transverse momentum distributions of
final state jets for signal with $\Lambda=5$ TeV for left panel and
signal with $\Lambda=M_{e_{8}}$ for right panel and also background
for both panels. The final state electron's normalized transverse
momentum distributions are the same as the plots of the final state
jets in Figure 4. It is seen from Figure 4 that $P_{T}>50$ GeV cut
for both final states essentially reduces background but signal is
almost unchanged. Left and right panels in Figure 5 represent normalized
pseudo-rapidity ($\eta$) distributions of electrons for signal with
$\Lambda=5$ TeV and signal with $\Lambda=M_{e_{8}}$ and also for
background, respectively. It is seen from left and right panels of
Figure 5, $\eta{}_{e^{-}}$ distributions of signal drastically different
from $\eta{}_{e^{-}}$ distributions of background. In addition, most
of background lie in $0<\eta_{e^{-}}<2$ region for both panels in
Figure 5. Left and right panels in Figure 6 present normalized pseudo-rapidity
($\eta$) distributions of jets for signal with $\Lambda=5$ TeV for
left panel and signal with $\Lambda=M_{e_{8}}$ for right panel and
also for background (both panels), respectively. In Figure 6, $\eta_{j}$
distributions for signal and background are not drastically different.
For this reason, we select pseudo-rapidity cuts values as follows:
$-4<\eta_{e^{-}}<-0.8$ for final state electron and $-4<\eta_{j}<1.5$
for final state jets. We present the invariant mass distributions
for signal with $\Lambda=5$ TeV for left panel and signal $\Lambda=M_{e_{8}}$
right panel and also for background (both panels) in Figure 7. 

\begin{figure}
\includegraphics[scale=0.6]{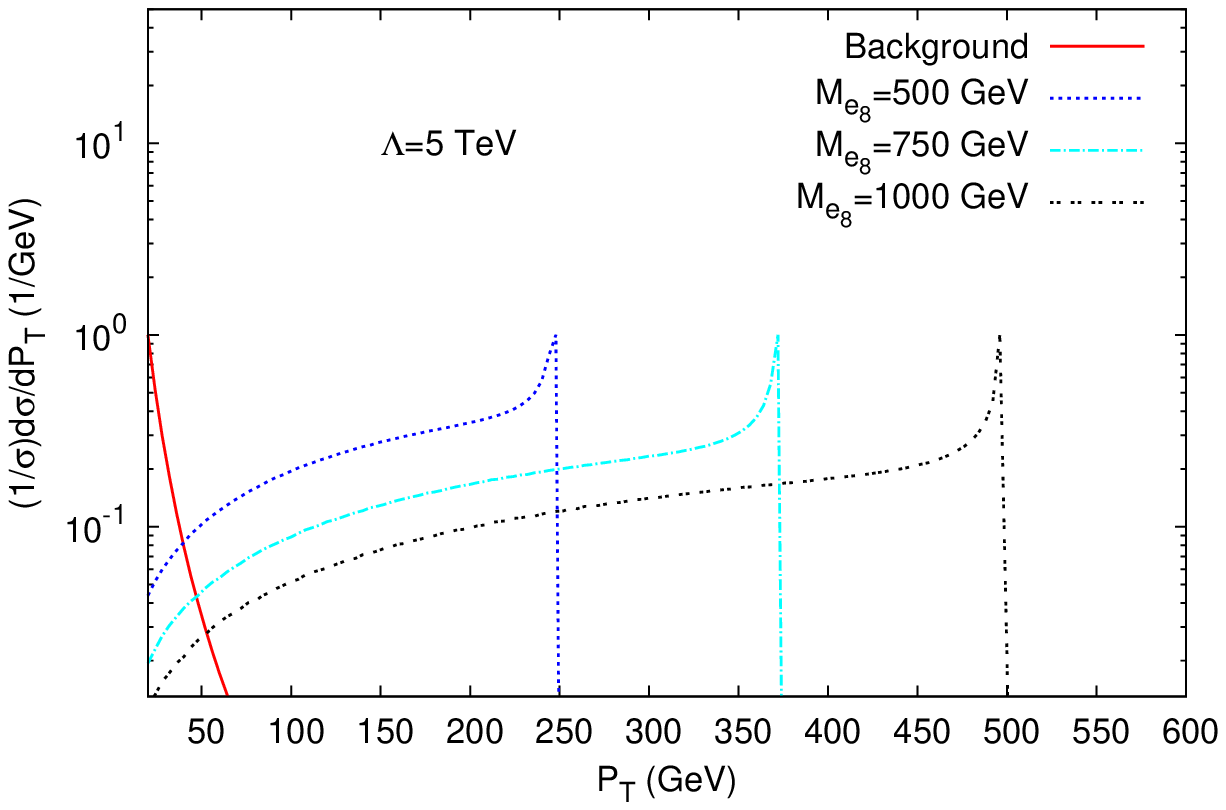}\includegraphics[scale=0.6]{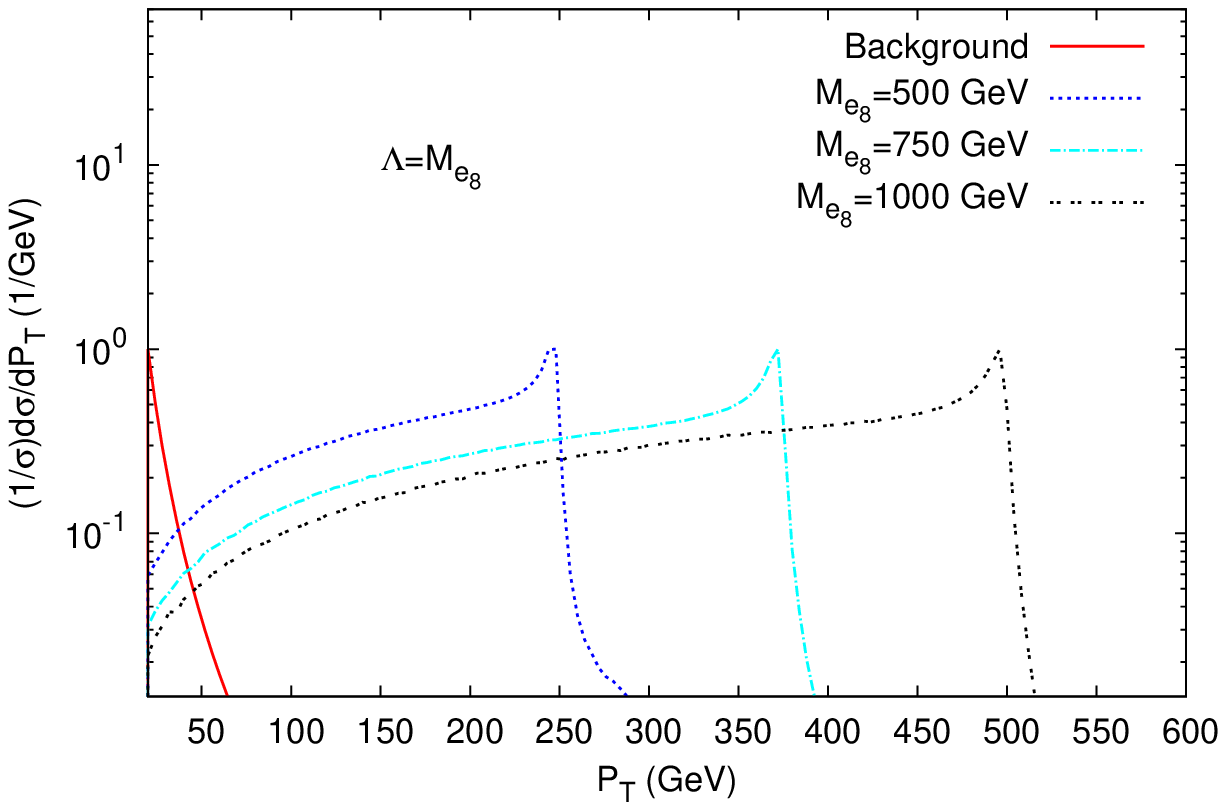}

\caption{Left: Normalized transverse momentum ($P_{T}$) distributions of signal
with $\Lambda=5$ TeV and background at the LHeC-1 with $\sqrt{s}=1.296$
TeV. Right: Normalized transverse momentum ($P_{T}$) distributions
of signal with $\Lambda=M_{e_{8}}$ TeV and background at the LHeC-1
with $\sqrt{s}=1.296$ TeV. }
\end{figure}

\begin{figure}
\includegraphics[scale=0.6]{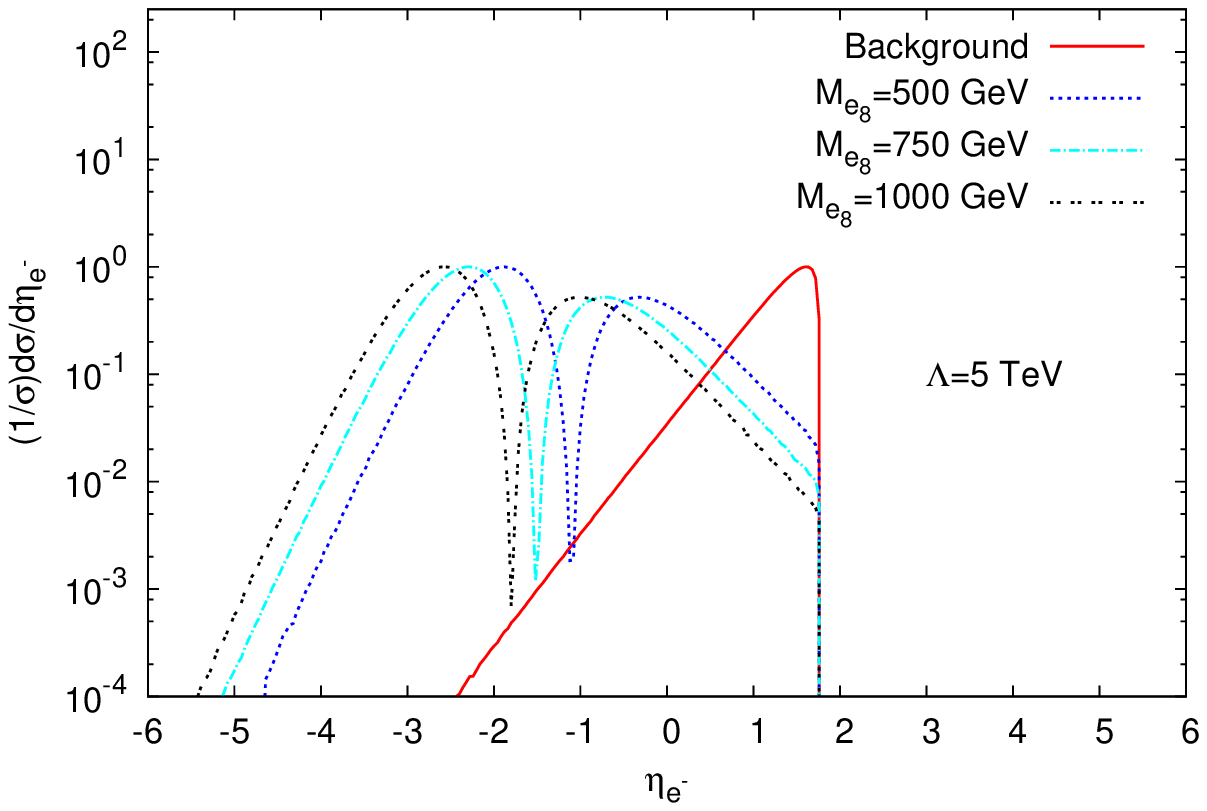}\includegraphics[scale=0.6]{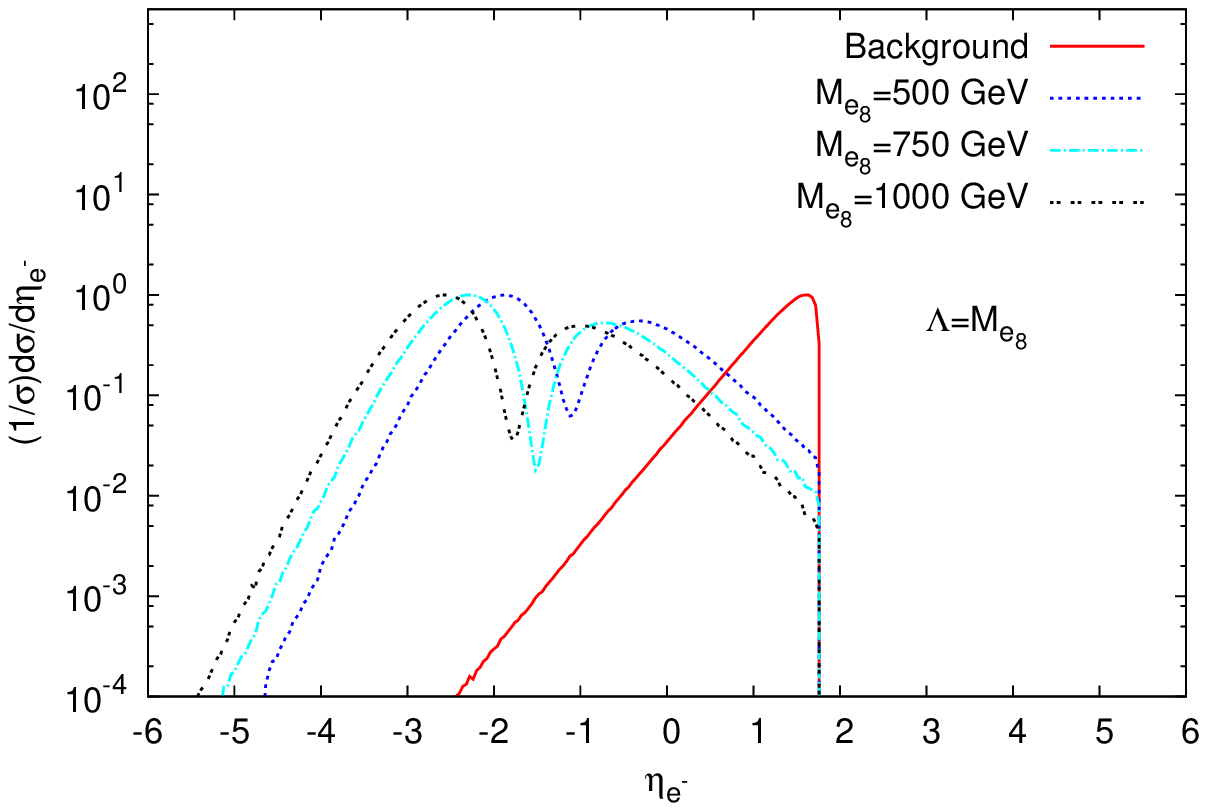}

\caption{Left: Normalized pseudo-rapidity distributions of final state electrons
for signal with $\Lambda=5$ TeV and background at the LHeC-1 with
$\sqrt{s}=1.296$ TeV. Right: Normalized pseudo-rapidity distributions
of final state electrons for signal with $\Lambda=M_{e_{8}}$ and
background at the LHeC-1 with $\sqrt{s}=1.296$ TeV.}
\end{figure}

\begin{figure}
\includegraphics[scale=0.6]{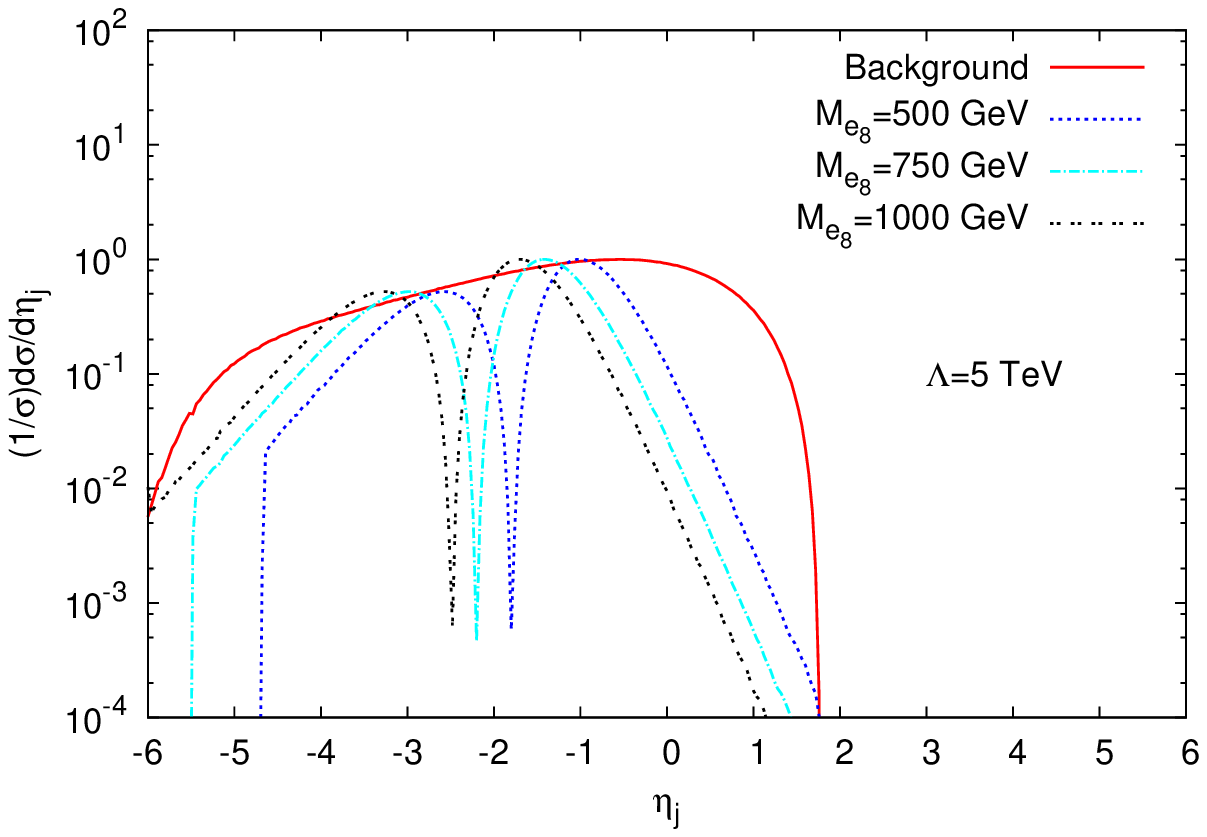}\includegraphics[scale=0.6]{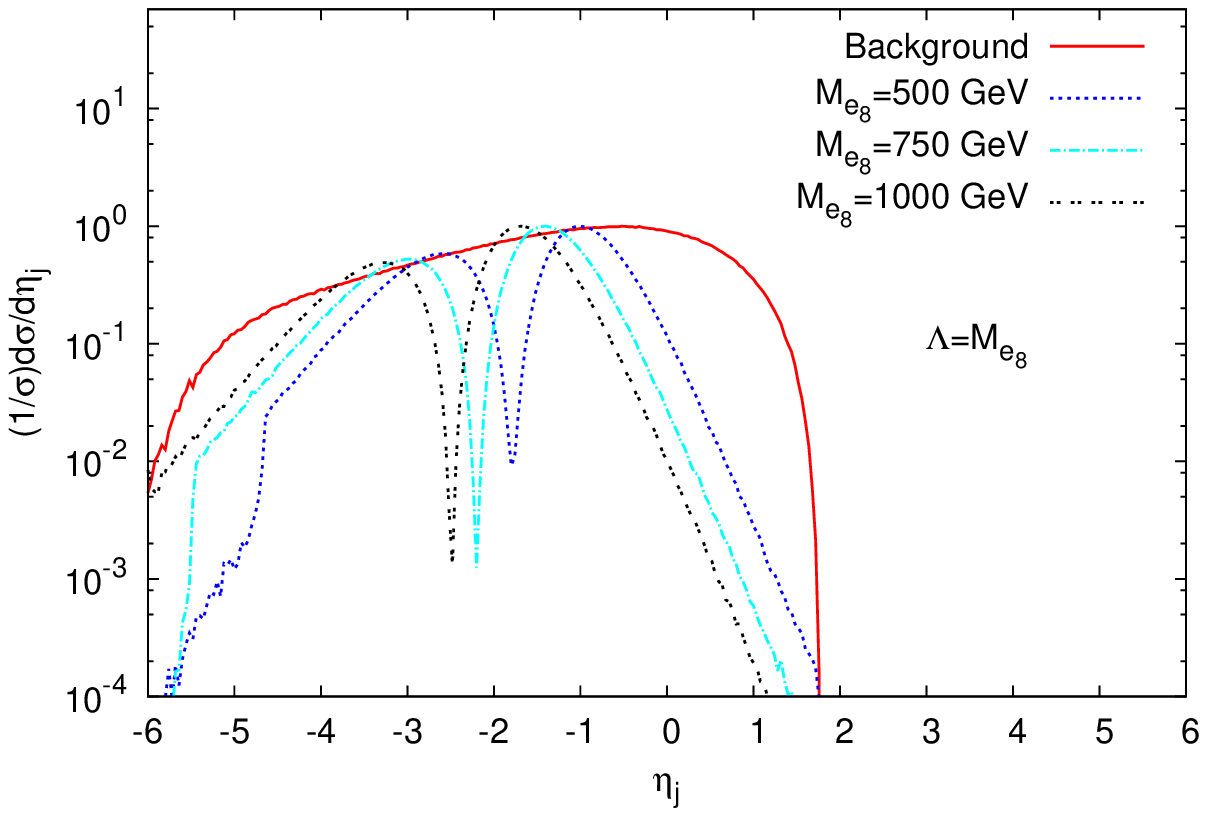}

\caption{Left: Normalized pseudo-rapidity distributions of final state jets
for signal with $\Lambda=5$ TeV and background at the LHeC-1 with
$\sqrt{s}=1.296$ TeV. Right: Normalized pseudo-rapidity distributions
of final state jets for signal with $\Lambda=M_{e_{8}}$ TeV and background
at the LHeC-1 with $\sqrt{s}=1.296$ TeV. }
\end{figure}

\begin{figure}
\includegraphics[scale=0.6]{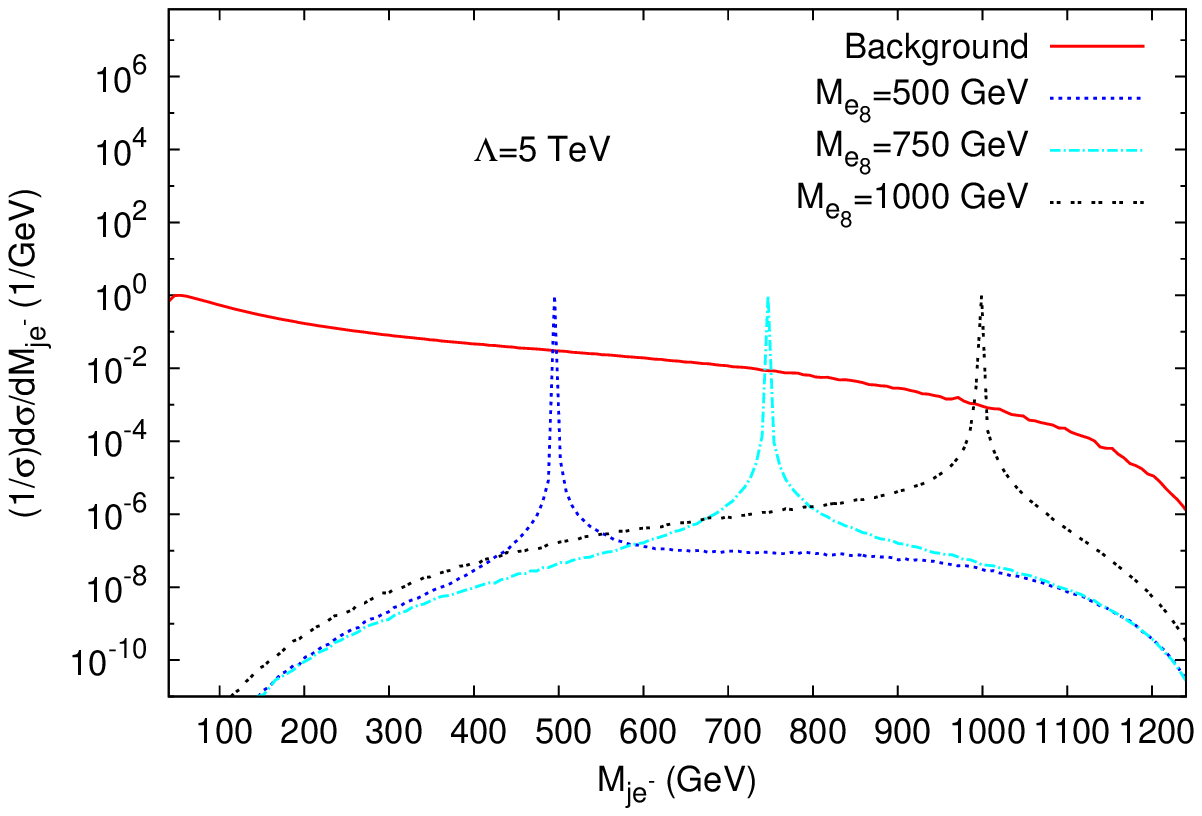}\includegraphics[scale=0.6]{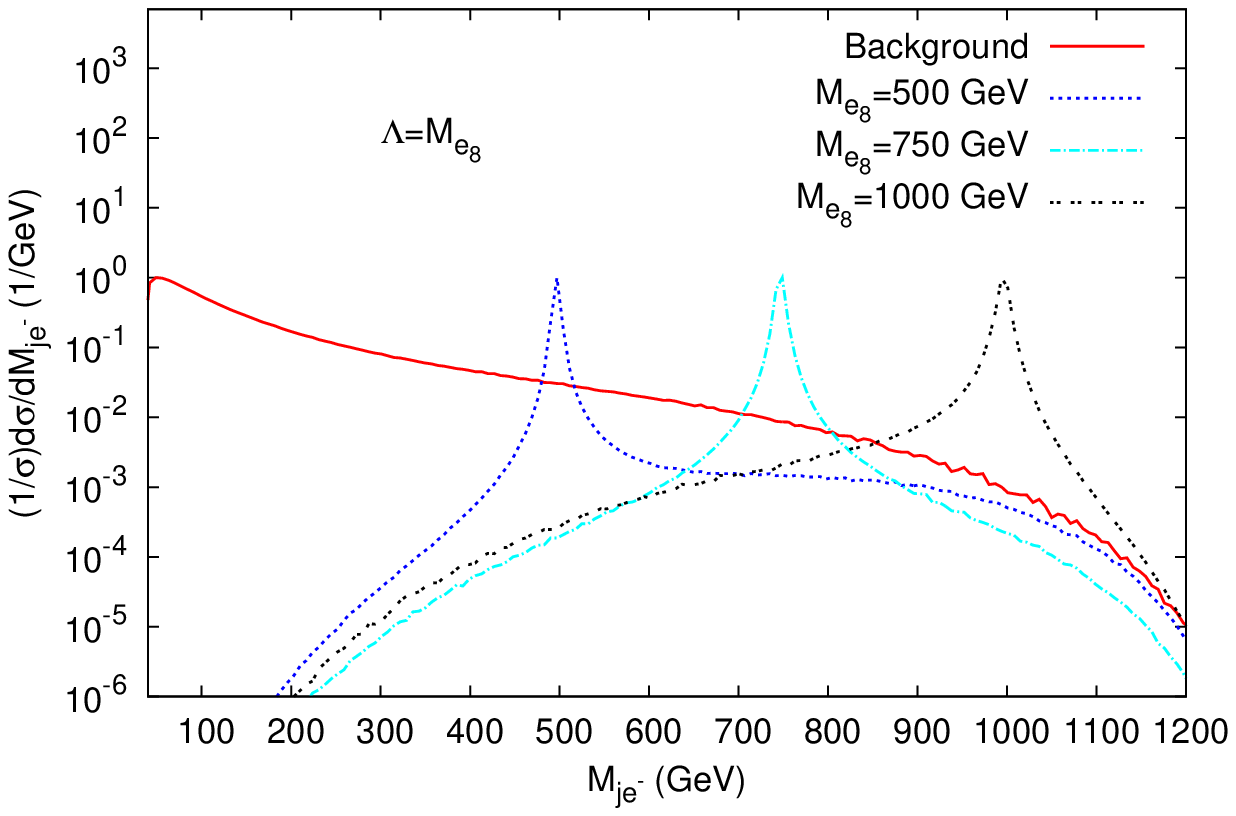}

\caption{Left: Normalized invariant mass distributions for signal with $\Lambda=5$
TeV and background at the LHeC-1 with $\sqrt{s}=1.296$ TeV. Right:
Normalized invariant mass distributions for signal with $\Lambda=M_{e_{8}}$
and background at the LHeC-1 with $\sqrt{s}=1.296$ TeV. }
\end{figure}

In order to extract the spin-3/2 color octet electron ($e_{8}$) signals
(for the $\Lambda=5$ TeV and $\Lambda=M_{e_{8}}$) and to suppress
the backgrounds we used $P_{T}>50$ GeV for all final state electron
and jets (these cut values are determined by Figure 4), $-4<\eta_{e^{-}}<-0.8$
for final state electron and $-4<\eta_{j}<1.5$ for final state jets
(these cut values are determined by Figure 5 and Figure 6). In addition
to these cut values, we used mass windows as $M_{e_{8}}-10$ GeV $<M_{je^{-}}<M_{e_{8}}+10$
GeV for signal with $\Lambda=5$ TeV and for background. It is seen
from Figure 1, signal decay width values are quite small when they
are compared to $10$ GeV for the $\Lambda=5$ TeV. We also used mass
window as $M_{e_{8}}-20$ GeV $<M_{je^{-}}<M_{e_{8}}+20$ GeV for
signal with $\Lambda=M_{e_{8}}$ and for background. As it seen from
Figure 1, signal decay width values approximately are $10$ GeV at
the LHeC-1 kinematic limit ($1296$ GeV) for the $\Lambda=M_{e_{8}}$. 

By using cut sets that are mentioned above and $L_{int}=10\, fb^{-1}$
integrated luminosity values of the LHeC-1, we have calculated event
numbers for some mass values of $e_{8}$ and background. We present
these event numbers in Table 2. 

\begin{table}
\begin{tabular}{|c|c|c|c|c|}
\hline 
$M_{e_{8}}$, GeV & \multicolumn{2}{c|}{$\Lambda=5$ TeV} & \multicolumn{2}{c|}{$\Lambda=M_{e_{8}}$}\tabularnewline
\hline 
\hline 
 & $N_{s}$ & $N_{b}$ & $N_{s}$ & $N_{b}$\tabularnewline
\hline 
$500$ & $304$ & $1268$ & $3559200$ & $2538$\tabularnewline
\hline 
$600$ & $185$ & $838$ & $1053100$ & $1678$\tabularnewline
\hline 
$700$ & $100$ & $514$ & $308510$ & $1028$\tabularnewline
\hline 
$800$ & $45$ & $279$ & $82767$ & $558$\tabularnewline
\hline 
$900$ & $16$ & $126$ & $18617$ & $252$\tabularnewline
\hline 
$1000$ & $4$ & $43$ & $3066$ & $86$\tabularnewline
\hline 
$1100$ & $-$ & $9$ & $278$ & $18$\tabularnewline
\hline 
$1200$ & $-$ & $-$ & $6$ & $1$\tabularnewline
\hline 
\end{tabular}

\caption{The number of signal and background events for the LHeC-1 with $L_{int}=10\, fb^{-1}$.
Here, $N_{s}$ is number of signal events and $N_{b}$ is number of
background events.}
\end{table}

For statistical significance, we use the following formula \cite{Bayatian}

\begin{equation}
S=\sqrt{2[(N_{s}+N_{b})ln(1+\frac{N_{s}}{N_{b}})-N_{s}]},
\end{equation}

where $N_{s}$ and $N_{b}$ represents the number of signal and background
events, respectively. The statistical significances of spin-3/2 color
octet electron signal with $\Lambda=5$ TeV are shown in Figure 8
for the LHeC-1 with $\sqrt{s}=1.296$ TeV and $L_{int}=10\, fb^{-1}$.

\begin{figure}
\includegraphics[scale=0.7]{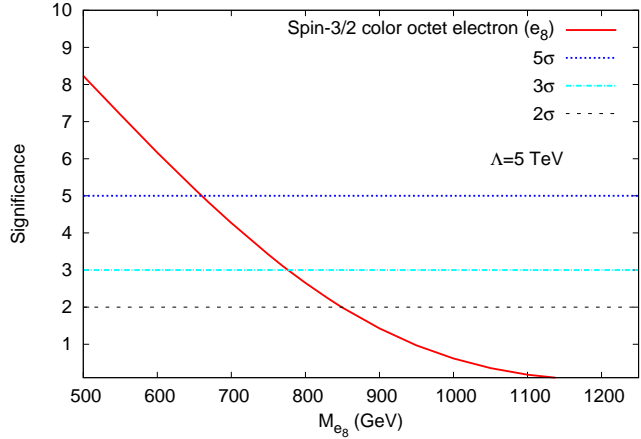}

\caption{The signal significances as a function of spin-3/2 color octet electron
mass at the LHeC-1 with $\sqrt{s}=1.296$ TeV. This figure is obtained
for $L_{int}=10\, fb^{-1}$ and $\Lambda=5$ TeV.}
\end{figure}

It is seen from Figure 8, upper mass limit for discovery ($5\sigma$)
of spin-3/2 color octet electron ($e_{8}$) is $660$ GeV at the LHeC-1
with $\sqrt{s}=1.296$ TeV. The upper observation ($3\sigma$) mass
limits of $e_{8}$ is $777$ GeV and the upper exclusion ($2\sigma$)
mass limits of $e_{8}$ is 849 GeV at the LHeC-1 with $\sqrt{s}=1.296$
TeV these values are obtained for $L_{int}=10\, fb^{-1}$and $\Lambda=5$
TeV. For the $\Lambda=M_{e_{8}}$ and $L_{int}=10\, fb^{-1}$, the
statistical significances of spin-3/2 color octet electron signal
at the LHeC-1 with $\sqrt{s}=1.296$ TeV are shown in Figure 9. It
is seen from Figure 9, upper mass limit for discovery ($5\sigma$)
of spin-3/2 color octet electron ($e_{8}$) is $1.19$ TeV at the
LHeC-1 with $\sqrt{s}=1.296$ TeV . The upper observation ($3\sigma$)
mass limits of $e_{8}$ is $1.21$ TeV and the upper exclusion ($2\sigma$)
mass limits of $e_{8}$ is $1.22$ TeV at the LHeC-1 with $\sqrt{s}=1.296$
TeV. These values are obtained for the compositeness scale $\Lambda=M_{e_{8}}$
and the integrated luminosity $L_{int}=10\, fb^{-1}$. 

\begin{figure}
\includegraphics[scale=0.7]{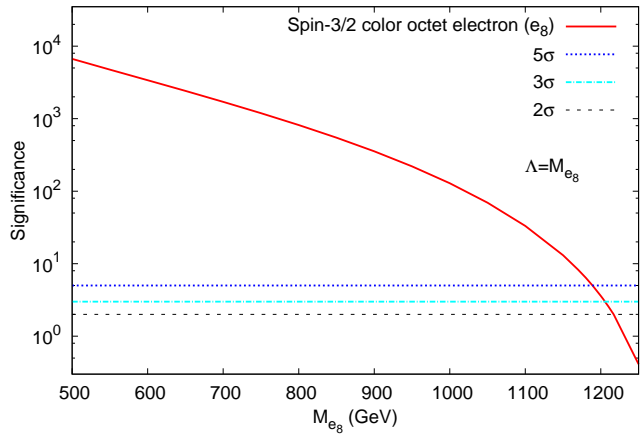}

\caption{The signal significances as a function of spin-3/2 color octet electron
mass at the LHeC-1 with $\sqrt{s}=1.296$ TeV. This figure is obtained
for $L_{int}=10\, fb^{-1}$ and $\Lambda=M_{e_{8}}$.}
\end{figure}

Spin-1/2 color octet electron decays into gluon and electron \cite{PDG,Celikel,Kantar,Hewett,Rizzo1,Rizzo2,Sahin,Akay,Mandal}
like as spin-3/2 color octet electron. Namely they will have same
final state in detector. In order to separate spin-3/2 color octet
electron signal from spin-1/2 color octet electron signal, we plot
normalized differential cross sections as a function of $cos\theta$
in Figure 10. The spin-1/2 color octet electron is produced mostly
in backward direction and its minimum cross section values are in
forward direction. However, spin-3/2 color octet electron is produced
mostly in both directions (forward and backward) and its minimum cross
section values are between in -0.4 to -0.3. Therefore, the spin-3/2
color octet electron shows more different angular shape than the spin-1/2
color octet electron.

\begin{figure}
\includegraphics[scale=0.6]{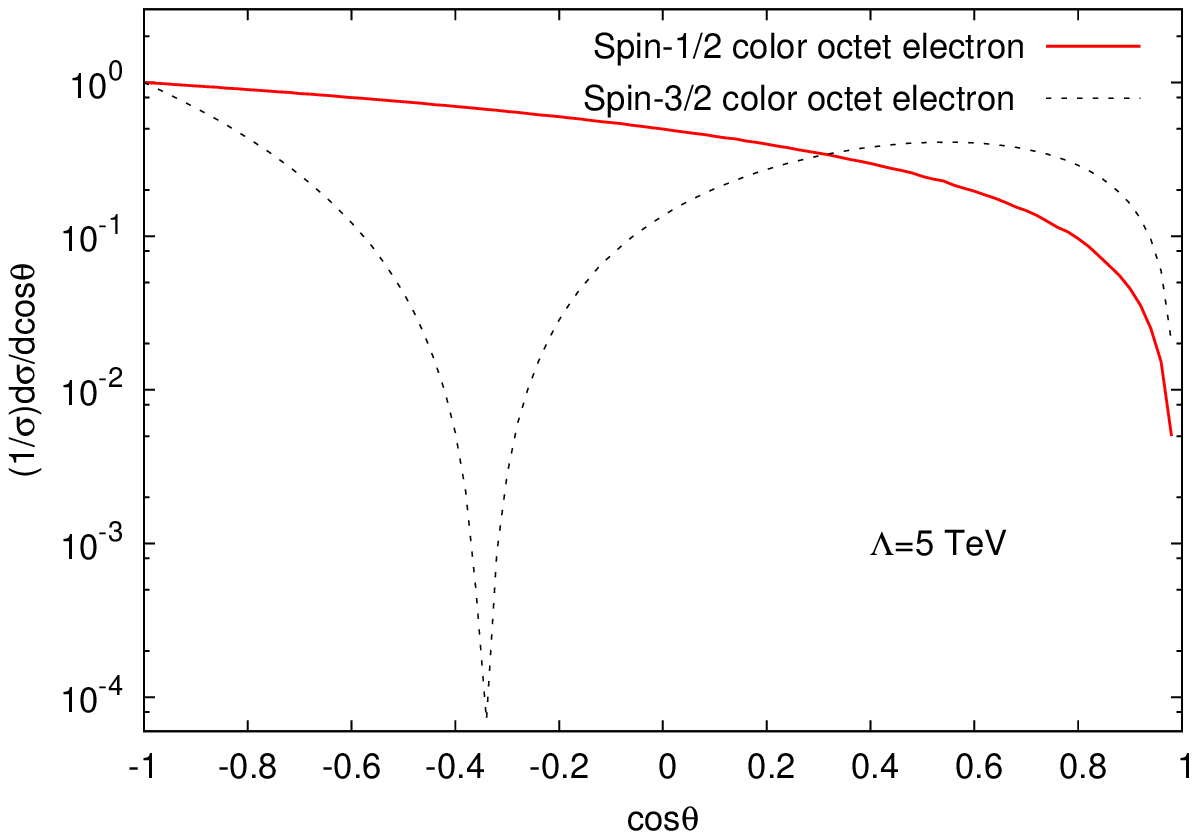}\includegraphics[scale=0.6]{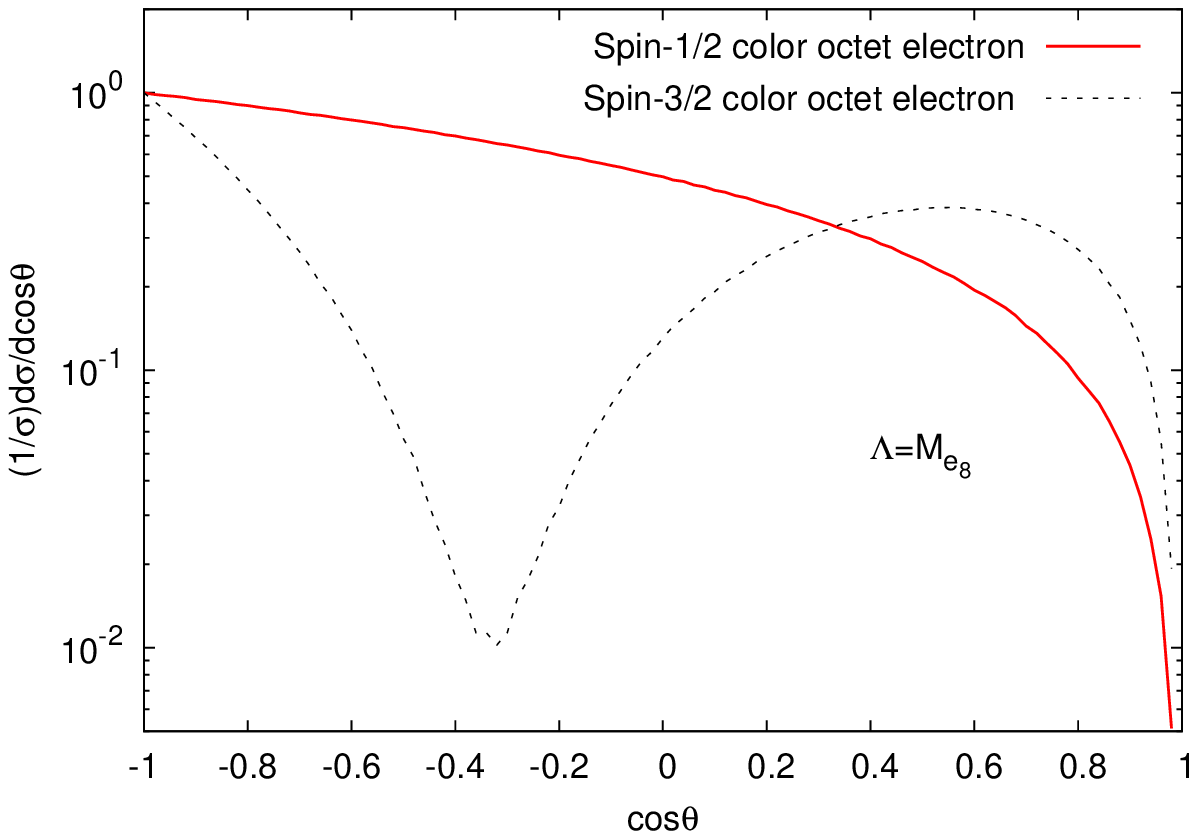}

\caption{Left:The differential cross section as a function the scattering angle
for the spin-3/2 color octet electron ($\Lambda=5$ TeV), and the
spin-1/2 color octet electron ($\Lambda=5$ TeV) at the LHeC-1 with
$\sqrt{s}=1.296$ TeV. Right: The differential cross section as a
function the scattering angle for the spin-3/2 color octet electron
($\Lambda=M_{e_{8}}$), and the spin-1/2 color octet electron ($\Lambda=M_{e_{8}}$)
at the LHeC-1 with $\sqrt{s}=1.296$ TeV. In the both panels, spin-1/2
and spin-3/2 color octet electron mass values are taken as $500$
GeV. }
\end{figure}

We have calculated the compositeness scale values for some spin-3/2
color octet masses at the LHeC-1 with $\sqrt{s}=1.296$ TeV and $L_{int}=10\, fb^{-1}$.
These values are presented as a function of spin-3/2 color octet electron
mass in Fig 11 and in Table 3. 

\begin{figure}
\includegraphics[scale=0.7]{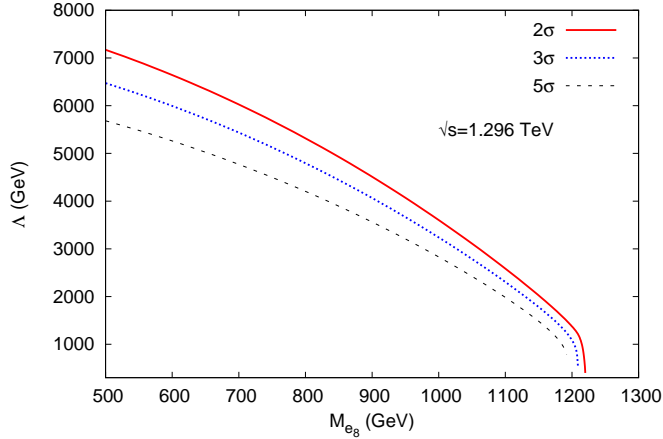}

\caption{Reachable values of the compositeness scale as a function of spin-3/2
color octet electron mass for the LHeC-1 with $\sqrt{s}=1.296$ TeV
and $L_{int}=10\, fb^{-1}$.}
\end{figure}

\begin{table}
\begin{tabular}{|c|c|c|c|}
\hline 
\multirow{2}{*}{$M_{e_{8}}$, GeV} & \multicolumn{3}{c|}{$\Lambda$, TeV}\tabularnewline
\cline{2-4} 
 & $5\sigma$ & $3\sigma$ & $2\sigma$\tabularnewline
\hline 
\hline 
$500$ & $5.68$ & $6.47$ & $7.17$\tabularnewline
\hline 
$600$ & $5.28$ & $6.01$ & $6.66$\tabularnewline
\hline 
$700$ & $4.80$ & $5.47$ & $6.07$\tabularnewline
\hline 
$800$ & $4.24$ & $4.85$ & $5.38$\tabularnewline
\hline 
$900$ & $3.61$ & $4.13$ & $4.59$\tabularnewline
\hline 
$1000$ & $2.89$ & $3.32$ & $3.69$\tabularnewline
\hline 
$1100$ & $2.05$ & $2.39$ & $2.68$\tabularnewline
\hline 
\end{tabular}

\caption{Reachable values of the compositeness scale for some spin-3/2 electrons
mass values at the LHeC-1 with $\sqrt{s}=1.296$ TeV and $L_{int}=10\, fb^{-1}$.}
\end{table}

It is seen from table 3 that the spin-3/2 color octet electron with
$M_{e_{8}}=500$ GeV can be discovered up to $5.68$ TeV compositeness
scale value at the LHeC-1 ($\sqrt{s}=1.296$ TeV and $L_{int}=10\, fb^{-1}$).

\subsection{Higher-energy LHeC ERL (LHeC-2 with $\sqrt{s}=2.049$ TeV and $L_{int}=1000\, fb^{-1}$)}

This LHeC option's the center of mass energy is $2.049$ TeV and integrated
luminosity is $L_{int}=1000\, fb^{-1}$. As mentioned in previous
subsection, our signal process is $ep\rightarrow e_{8}\rightarrow eg+X$
and background process is $ep\rightarrow ej+X$ through $\gamma$
and $Z$ exchange, where g represents gluon and $j$ represents jets
that are composed of quarks ($u,$ $\bar{u}$, $d,$ $\bar{d}$, $c$,
$\bar{c}$, $s$, $\bar{s}$, $b$, $\bar{b}$ ). In order to reduce
backgrounds and obtain clear signal, we need to apply some cuts. So,
transverse momentum ($P_{T}$ ) and pseudo-rapidity ($\eta$) distributions
of signal and background processes are used for determining appropriate
cuts. Figure 12 presents normalized transverse momentum distributions
of final state jets for signal with $\Lambda=5$ TeV for left panel,
signal with $\Lambda=M_{e_{8}}$ for right panel and also for background
(both panels). Normalized transverse momentum distributions of final
state electron for signal with $\Lambda=5$ TeV and signal with $\Lambda=M_{e_{8}}$
and also for background are the same as normalized transverse momentum
distributions of the final state jets in Figure 12. It is seen from
left and right panels of Figure 12, $P_{T}>50$ GeV cuts for the final
state electron and jets essentially reduces background but signal
is almost unchanged. Left and right panels in Figure 13 represent
normalized pseudo-rapidity ($\eta$) distributions of electron for
signal with $\Lambda=5$ TeV and signal with $\Lambda=M_{e_{8}}$
and also for background, respectively. As it seen from left and right
panels of Figure 13, $\eta{}_{e^{-}}$distributions of signal drastically
different from $\eta{}_{e^{-}}$ distributions of background. It is
seen that most of background lie in $0<\eta_{e^{-}}<3$ region in
both panels of Figure 13. Left and right panels in Figure 14 present
normalized pseudo-rapidity ($\eta$) distributions of jets for signal
with $\Lambda=5$ TeV for left panel, signal with $\Lambda=M_{e_{8}}$
for right panel and also for background (both panels), respectively.
As it is seen from Figure 14 in both left and right panels, $\eta_{j}$
distributions for signal and background are not drastically different.
Transverse momentum ($P_{T}$) and pseudo-rapidity distributions ($\eta$)
of signal with $\Lambda=10$ TeV give the same knowledge as transverse
and pseudo-rapidity distributions of signal with $\Lambda=5$ TeV
and $\Lambda=M_{e_{8}}$. Therefore, these distributions of signal
with $\Lambda=10$ TeV are not given in text. Taking advantage of
the this kinematical distributions, we select pseudo-rapidity cuts
values as follows: $-4<\eta_{e^{-}}<-0.3$ for final state electron
and $-4<\eta_{j}<2$ for final state jets. We present the invariant
mass distributions for signal with $\Lambda=5$ TeV for left panel,
signal $\Lambda=M_{e_{8}}$ for right panel and also for background
(both panels) in Figure 15. 

\begin{figure}
\includegraphics[scale=0.6]{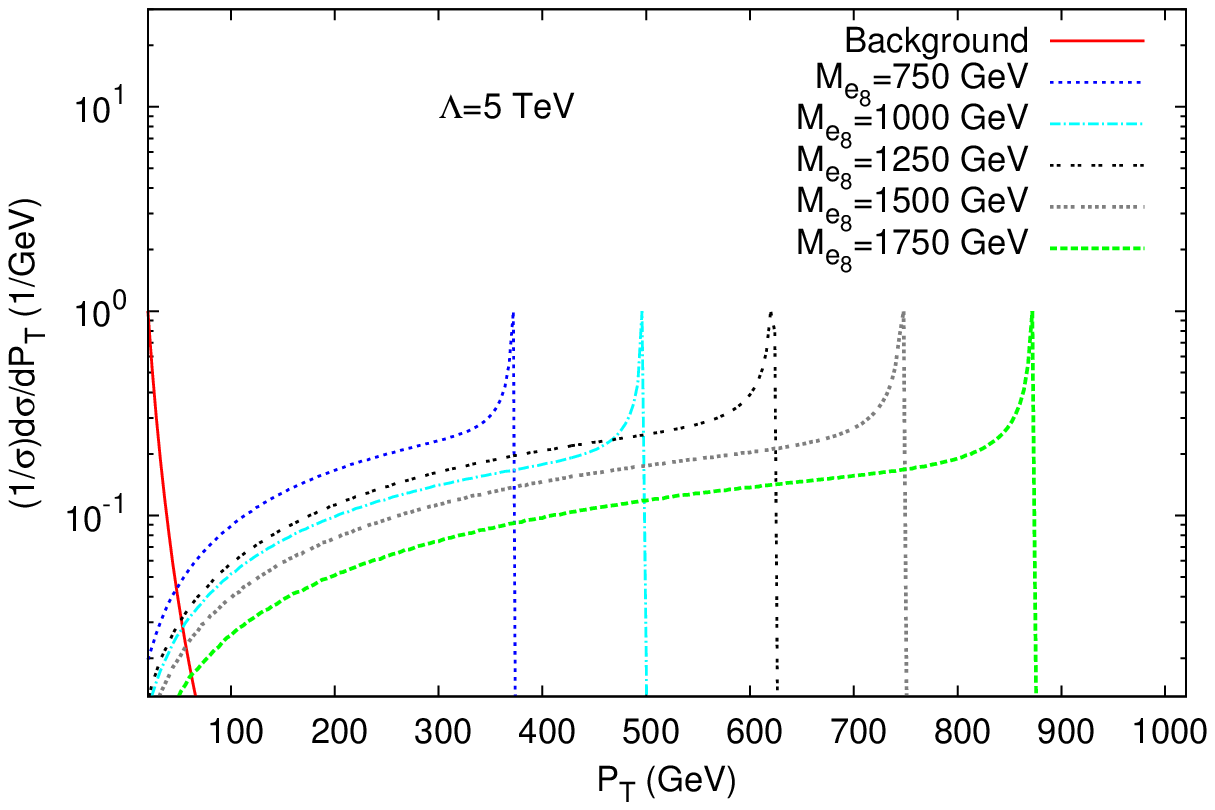}\includegraphics[scale=0.6]{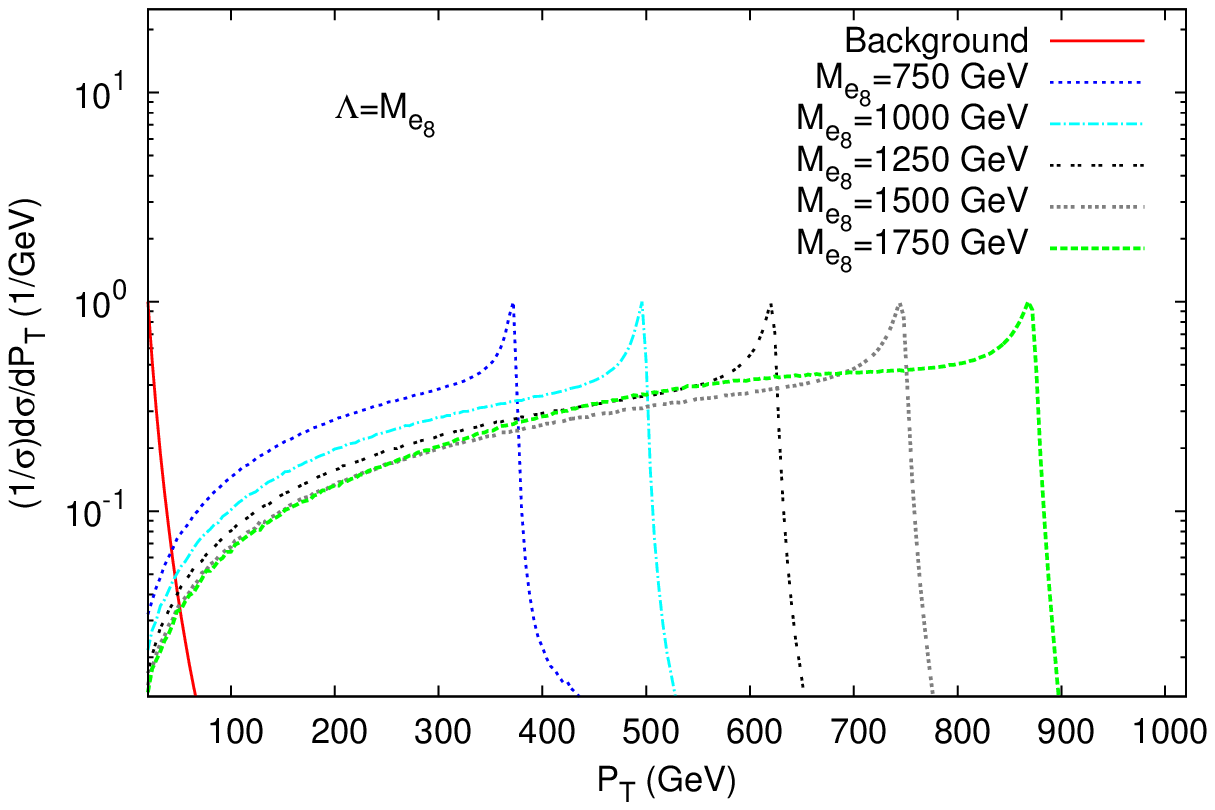}

\caption{Left: Normalized transverse momentum ($P_{T}$) distributions of the
final state jets for signal with $\Lambda=5$ TeV and for background
at the LHeC-2. Right: Normalized transverse momentum ($P_{T}$) distributions
of the final state jets for signal with $\Lambda=M_{e_{8}}$ TeV and
for background at the LHeC-2. }
\end{figure}

\begin{figure}
\includegraphics[scale=0.6]{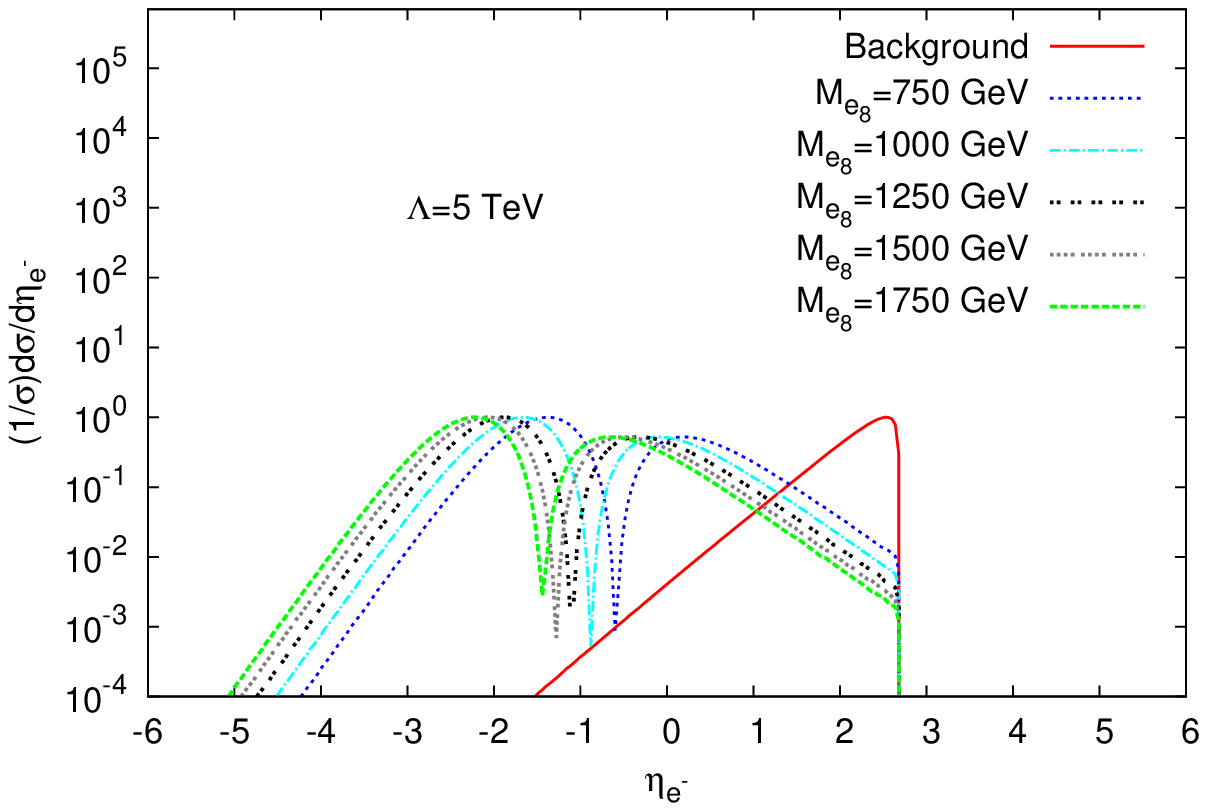}\includegraphics[scale=0.6]{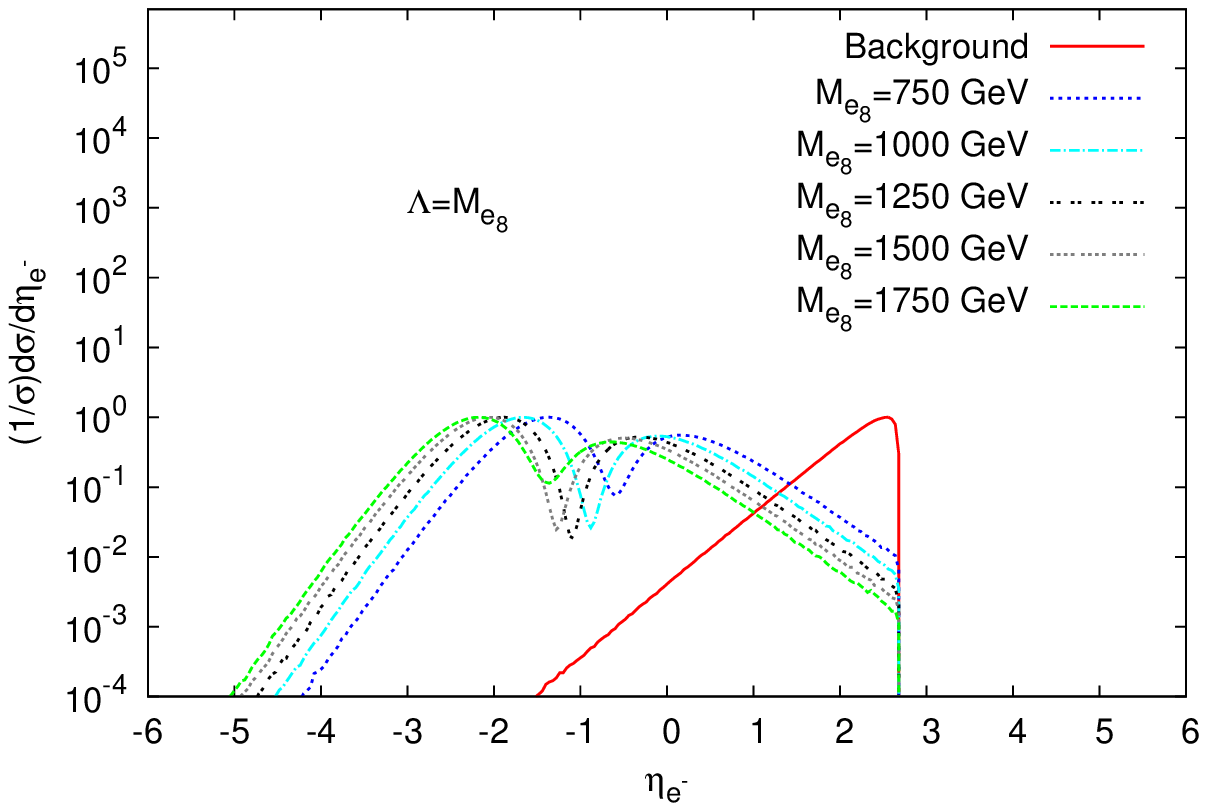}

\caption{Left: Normalized pseudo-rapidity distributions of final state electrons
for signal with $\Lambda=5$ TeV and background at the LHeC-2. Right:
Normalized pseudo-rapidity distributions of final state electrons
for signal with $\Lambda=M_{e_{8}}$ and background at the LHeC-2.}
\end{figure}

\begin{figure}
\includegraphics[scale=0.6]{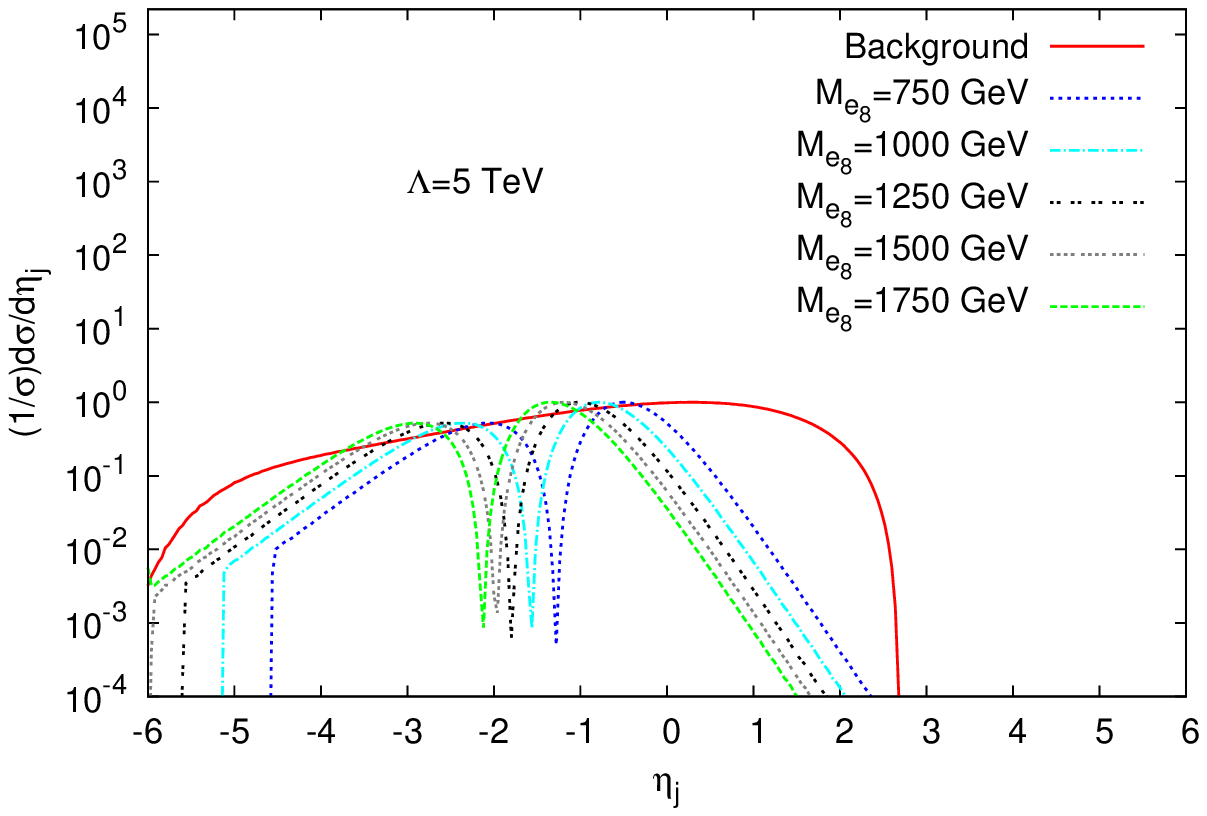}\includegraphics[scale=0.6]{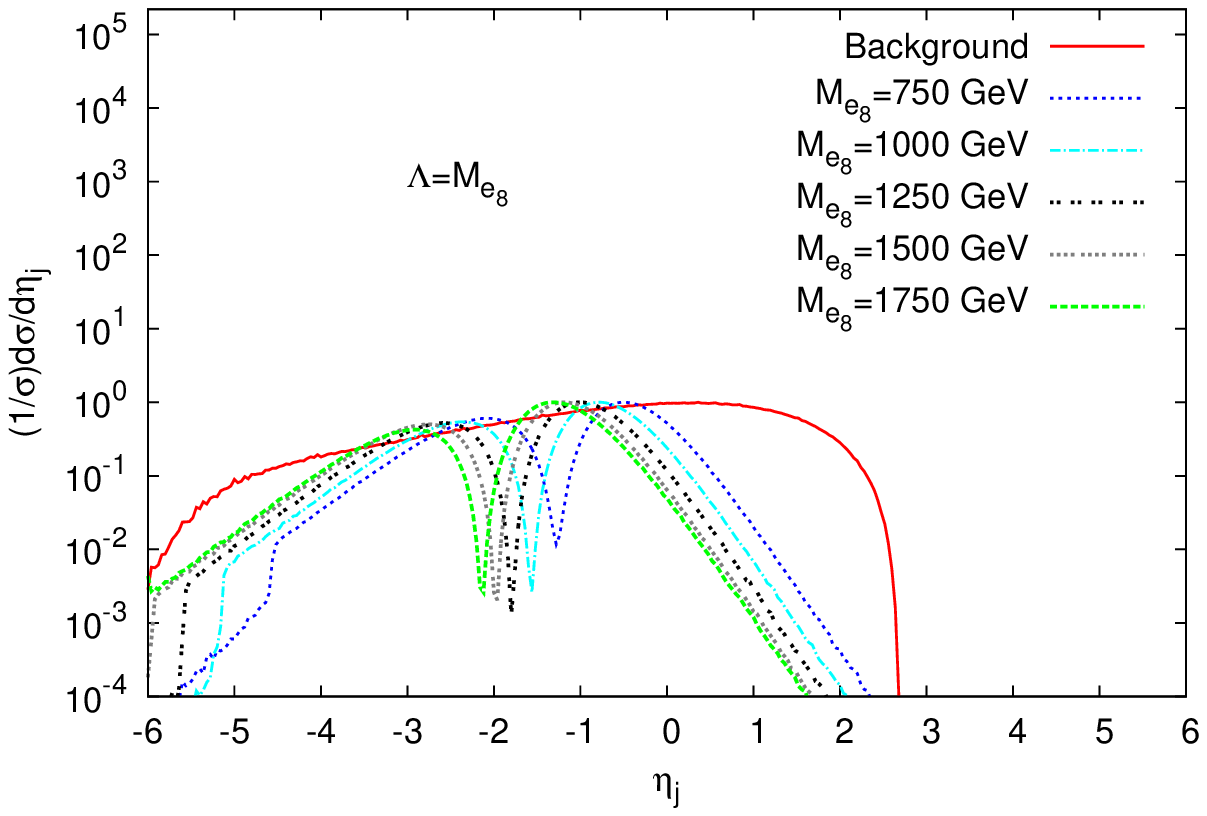}

\caption{Left: Normalized pseudo-rapidity distributions of final state jets
for signal with $\Lambda=5$ TeV and background at the LHeC-2. Right:
Normalized pseudo-rapidity distributions of final state jets for signal
with $\Lambda=M_{e_{8}}$ TeV and background at the LHeC-2. }
\end{figure}

\begin{figure}
\includegraphics[scale=0.6]{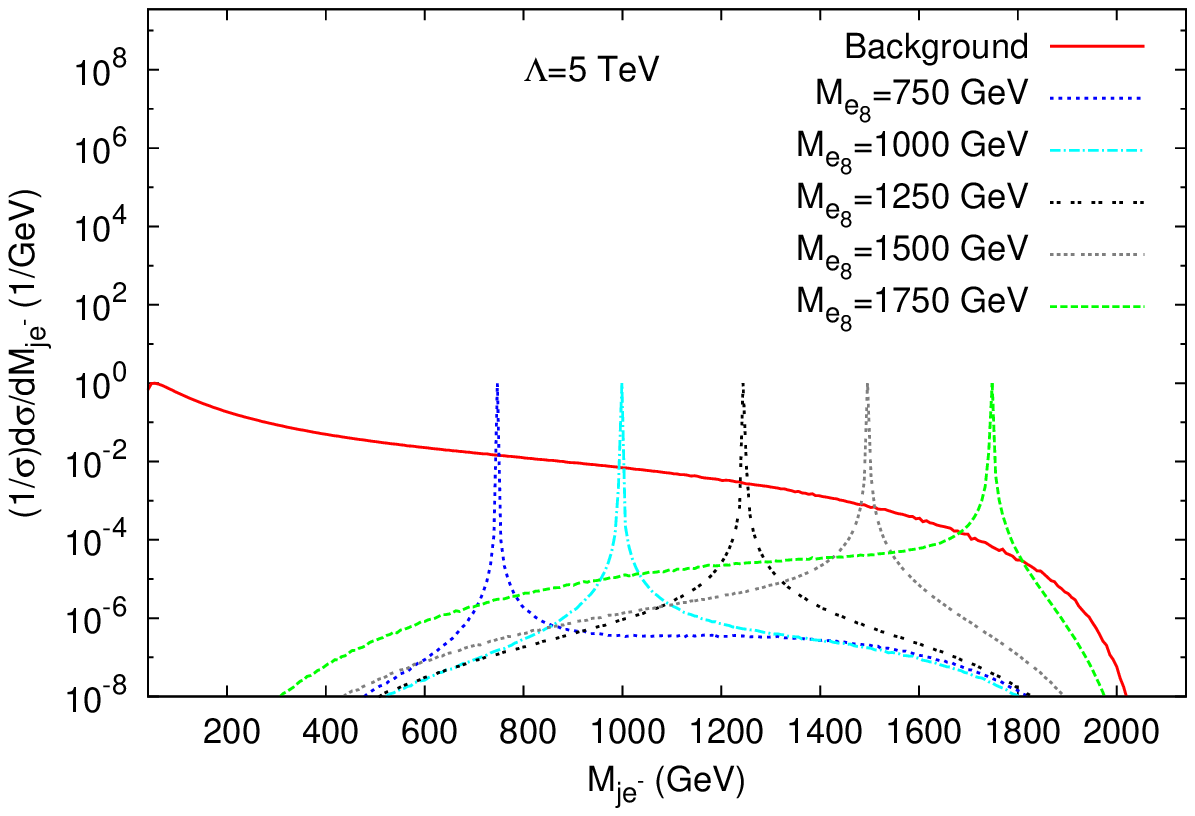}\includegraphics[scale=0.6]{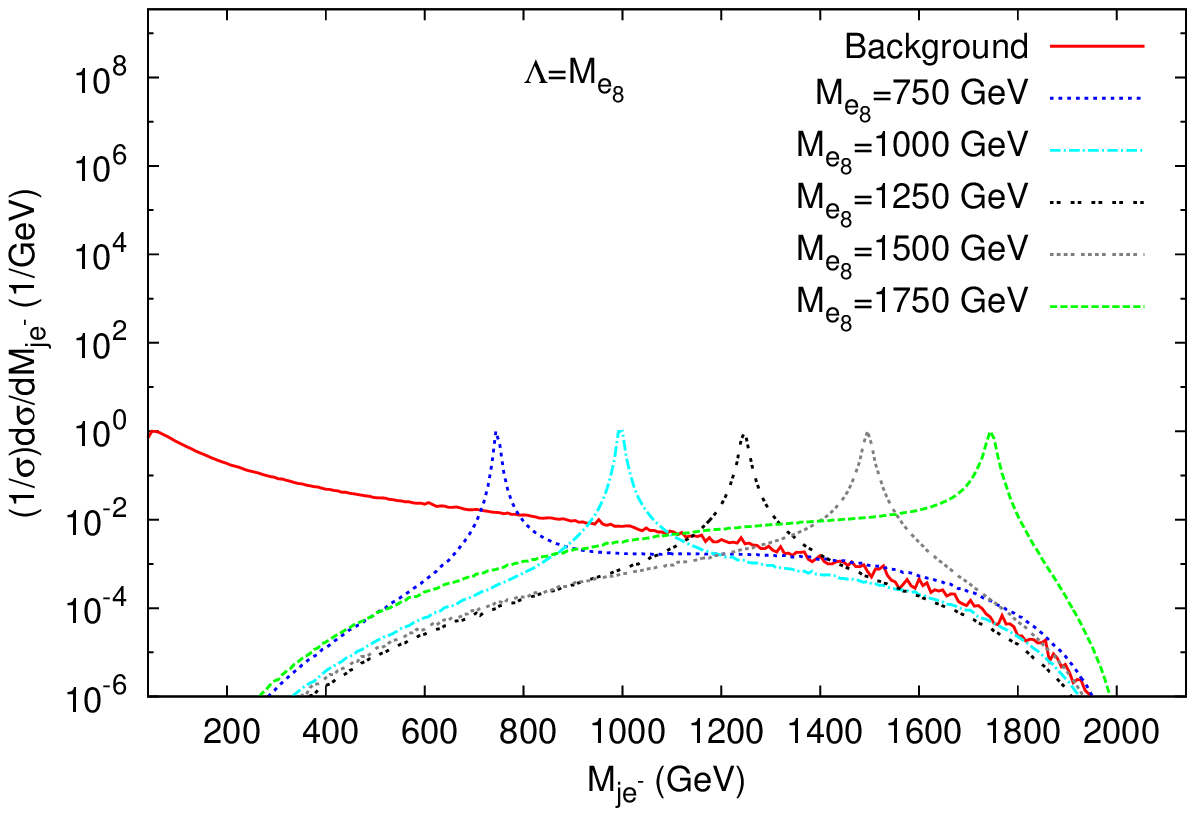}

\caption{Left: Normalized invariant mass distributions for signal with $\Lambda=5$
TeV and background at the LHeC-2. Right: Normalized invariant mass
distributions for signal with $\Lambda=M_{e_{8}}$ and background
at the LHeC-2. }
\end{figure}

In order to extract the spin-3/2 color octet electron ($e_{8}$) signals
for the $\Lambda=10$ TeV, $\Lambda=5$ TeV and $\Lambda=M_{e_{8}}$
and to suppress the backgrounds we used $P_{T}>50$ GeV for all final
state electron and jets (these cut values are determined by Figure
12), $-4<\eta_{e^{-}}<-0.3$ for final state electron and $-4<\eta_{j}<2$
for final state jets (these cut values are determined by Figure 13
and Figure 14). In addition to these cut values, we have used mass
window as $M_{e_{8}}-10$ GeV $<M_{je^{-}}<M_{e_{8}}+10$ GeV for
signal with $\Lambda=10$ TeV, $\Lambda=5$ TeV and also for background.
In view of the fact that it is seen from Figure 1, signal decay width
values are much more smaller than $10$ GeV for the $\Lambda=10$
TeV and $\Lambda=5$ TeV. In addition, we have used mass window as
$M_{e_{8}}-40$ GeV $<M_{je^{-}}<M_{e_{8}}+40$ GeV for signal with
$\Lambda=M_{e_{8}}$ and for background. Since, it is seen from Figure
1, signal decay width values are approximately $20$ GeV at the LHeC-2
kinematic limit ($2.049$ TeV) for the $\Lambda=M_{e_{8}}$. By using
this cut sets and integrated luminosity values of the LHeC-2, $L_{int}=1000\, fb^{-1}$,
we have calculated event numbers of some mass values of $e_{8}$ and
the background. We present these event numbers in Table 4. 

\begin{table}
\begin{tabular}{|c|c|c|c|c|c|c|}
\hline 
$M_{e_{8}}$, GeV & \multicolumn{2}{c|}{$\Lambda=10$ TeV} & \multicolumn{2}{c|}{$\Lambda=5$ TeV} & \multicolumn{2}{c|}{$\Lambda=M_{e_{8}}$}\tabularnewline
\hline 
\hline 
 & $N_{s}$ & $N{}_{b}$ & $N_{s}$ & $N{}_{b}$ & $N_{s}$ & $N{}_{b}$\tabularnewline
\hline 
$500$ & $8.3\times10^{3}$ & $7.74\times10^{4}$ & $1.33\times10^{5}$ & $7.75\times10^{4}$ & $1.60\times10^{9}$ & $3.11\times10^{5}$\tabularnewline
\hline 
$750$ & $4.61\times10^{3}$ & $3.96\times10^{4}$ & $7.38\times10^{4}$ & $3.96\times10^{4}$ & $1.82\times10^{8}$ & $1.57\times10^{5}$\tabularnewline
\hline 
$1000$ & $2.08\times10^{3}$ & $1.20\times10^{4}$ & $3.34\times10^{4}$ & $1.99\times10^{4}$ & $2.66\times10^{7}$ & $7.99\times10^{4}$\tabularnewline
\hline 
$1250$ & $6.70\times10^{2}$ & $8.05\times10^{3}$ & $1.07\times10^{4}$ & $8.05\times10^{3}$ & $3.53\times10^{6}$ & $3.18\times10^{4}$\tabularnewline
\hline 
$1500$ & $1.19\times10^{2}$ & $2.07\times10^{3}$ & $1.89\times10^{3}$ & $2.07\times10^{3}$ & $3.04\times10^{5}$ & $8.33\times10^{3}$\tabularnewline
\hline 
$1750$ & $6$ & $1.99\times10^{2}$ & $9.50\times10^{1}$ & $1.99\times10^{2}$ & $8.40\times10^{3}$ & $8.07\times10^{2}$\tabularnewline
\hline 
$2000$ & - & - & - & - & $1$ & -\tabularnewline
\hline 
\end{tabular}

\caption{The number of signal and background event for the LHeC-2 with $L_{int}=1000\, fb^{-1}$.
$N_{s}$ and $N_{b}$ represent number of signal and background event,
respectively. }
\end{table}

For statistical significance, we have used the Equation 5 and we have
presented the statistical significances of spin-3/2 color octet electron
signal with $\Lambda=10$ TeV, $\Lambda=5$ TeV and $\Lambda=M_{e_{8}}$
as a function of spin-3/2 color octet electron masses in Figure 16,
17 and 18. Figure 16, 17 and 18 are obtained for LHeC-2 with $\sqrt{s}=2.049$
TeV and $L_{int}=1000\, fb^{-1}$. 

\begin{figure}
\includegraphics[scale=0.7]{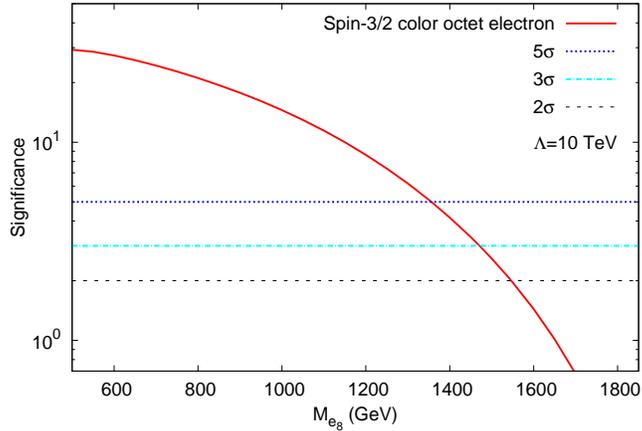}

\caption{The signal significances as a function of spin-3/2 color octet electron
mass at the LHeC-2 with $\sqrt{s}=2.049$ TeV. This figure is obtained
for $L_{int}=1000\, fb^{-1}$ and $\Lambda=10$ TeV.}
\end{figure}

As one can see from Figure 16, the upper mass limit for discovery
($5\sigma$) of spin-3/2 color octet electron ($e_{8}$) is $1.36$
TeV at the LHeC-2 with $\sqrt{s}=2.049$ TeV. The upper observation
($3\sigma$) mass limit of $e{}_{8}$ is $1.47$ TeV and the upper
exclusion limit of $e_{8}$ is $1.55$ TeV. These upper mass values
are obtained for $L_{int}=1000\, fb^{-1}$ integrated luminosity and
$\Lambda=10$ TeV.

\begin{figure}
\includegraphics[scale=0.7]{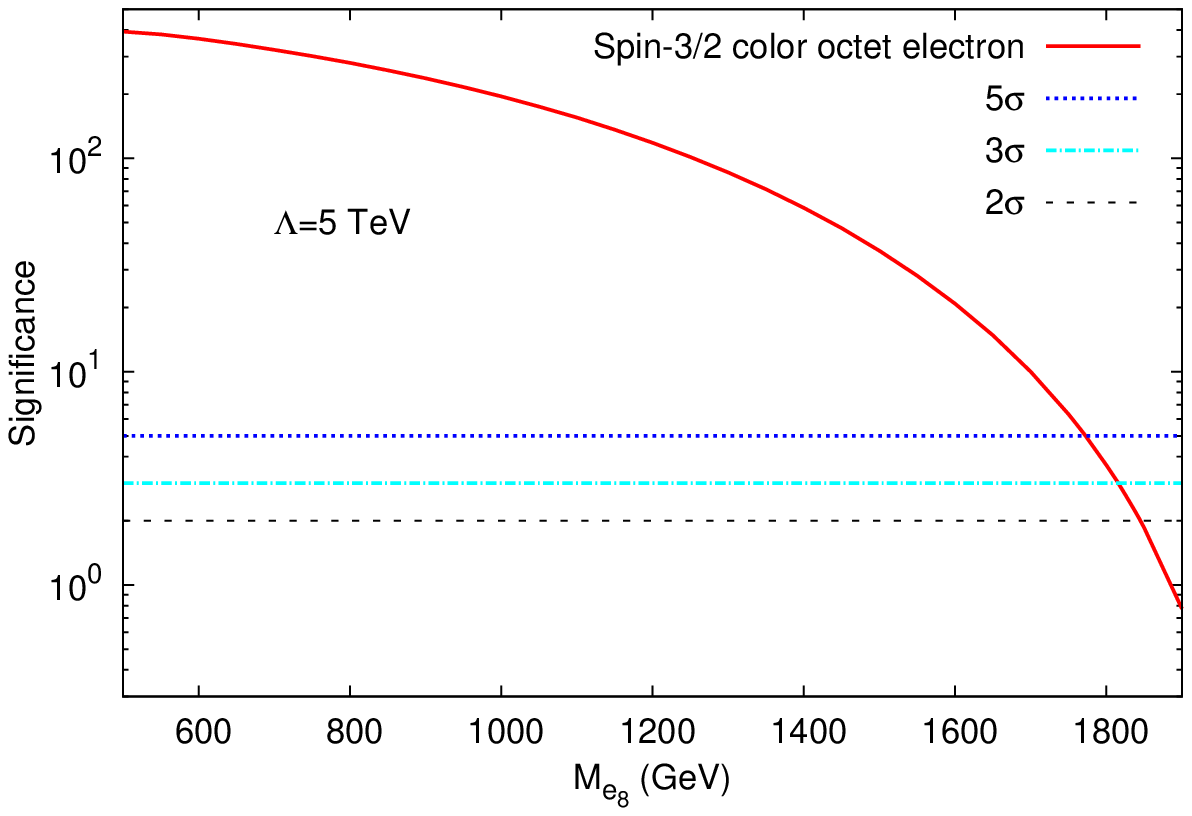}

\caption{The signal significances as a function of spin-3/2 color octet electron
mass at the LHeC-2 with $\sqrt{s}=2.049$ TeV. This figure is obtained
for $L_{int}=1000\, fb^{-1}$ and $\Lambda=5$ TeV.}
\end{figure}

It can be seen from Figure 17, the upper mass limit for discovery
($5\sigma$) of spin-3/2 color octet electron ($e_{8}$) is $1.77$
TeV at the LHeC-2 with $\sqrt{s}=2.049$ TeV. The upper observation
($3\sigma$) mass limit of $e{}_{8}$ is $1.82$ TeV and the upper
exclusion limit of $e_{8}$ is $1.85$ TeV. These upper mass values
are obtained for $L_{int}=1000\, fb^{-1}$ integrated luminosity and
$\Lambda=5$ TeV.

\begin{figure}
\includegraphics[scale=0.7]{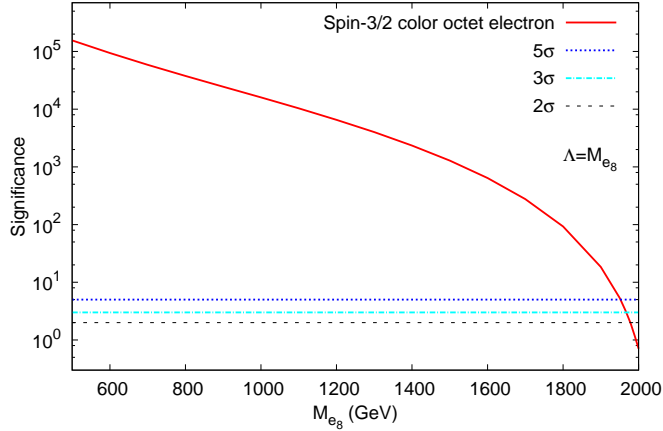}

\caption{The signal significances as a function of spin-3/2 color octet electron
mass at the LHeC-2 with $\sqrt{s}=2.049$ TeV. This figure is obtained
for $L_{int}=1000\, fb^{-1}$ and $\Lambda=M_{e_{8}}$ TeV.}
\end{figure}

It is seen from Figure 18, the upper mass limit for discovery ($5\sigma$)
of spin-3/2 color octet electron ($e_{8}$) is $1.95$ TeV at the
LHeC-2 with $\sqrt{s}=2.049$ TeV. The upper observation ($3\sigma$)
mass limit of $e{}_{8}$ is $1.97$ TeV and the upper exclusion limit
of $e_{8}$ is $1.98$ TeV. These upper mass values are obtained for
$L_{int}=1000\, fb^{-1}$ integrated luminosity and $\Lambda=M_{e_{8}}$
TeV. 

The spin-1/2 color octet electron and the spin-3/2 color octet electron
will have the same final state at the LHeC-2 with $\sqrt{s}=2.049$
TeV. In order to diverge the spin-3/2 and the spin-1/2 color octet
electron signals we plot normalized differential cross section as
a function of $cos\theta$ in Figure 19 and Figure 20. The spin-1/2
color octet electron is produced mostly in backward direction and
it has its minimum cross section values in forward direction. Whereas,
spin-3/2 color octet electron is produced mostly in both directions
(forward and backward) and its minimum cross section values are between
in -0.4 to -0.3. Therefore, the spin-3/2 color octet electron shows
different angular shape from the spin-1/2 color octet electron.

\begin{figure}
\includegraphics[scale=0.6]{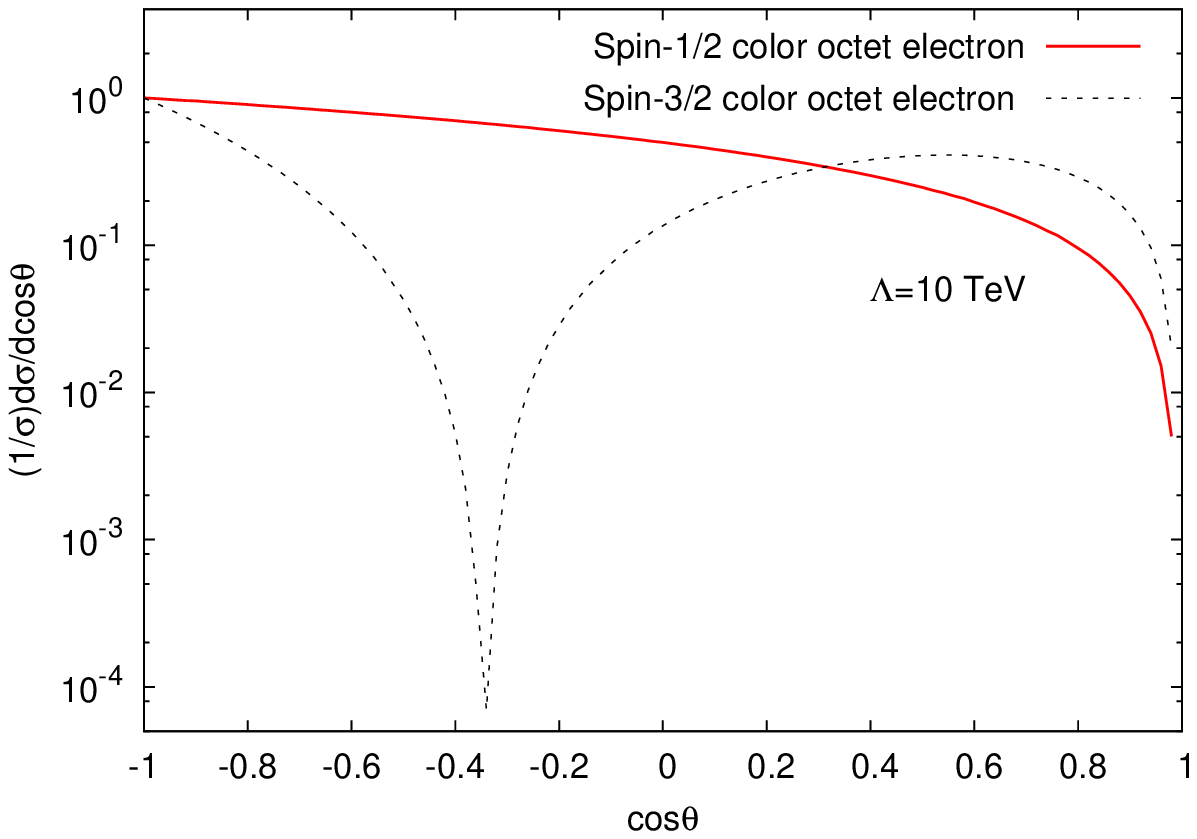}\includegraphics[scale=0.6]{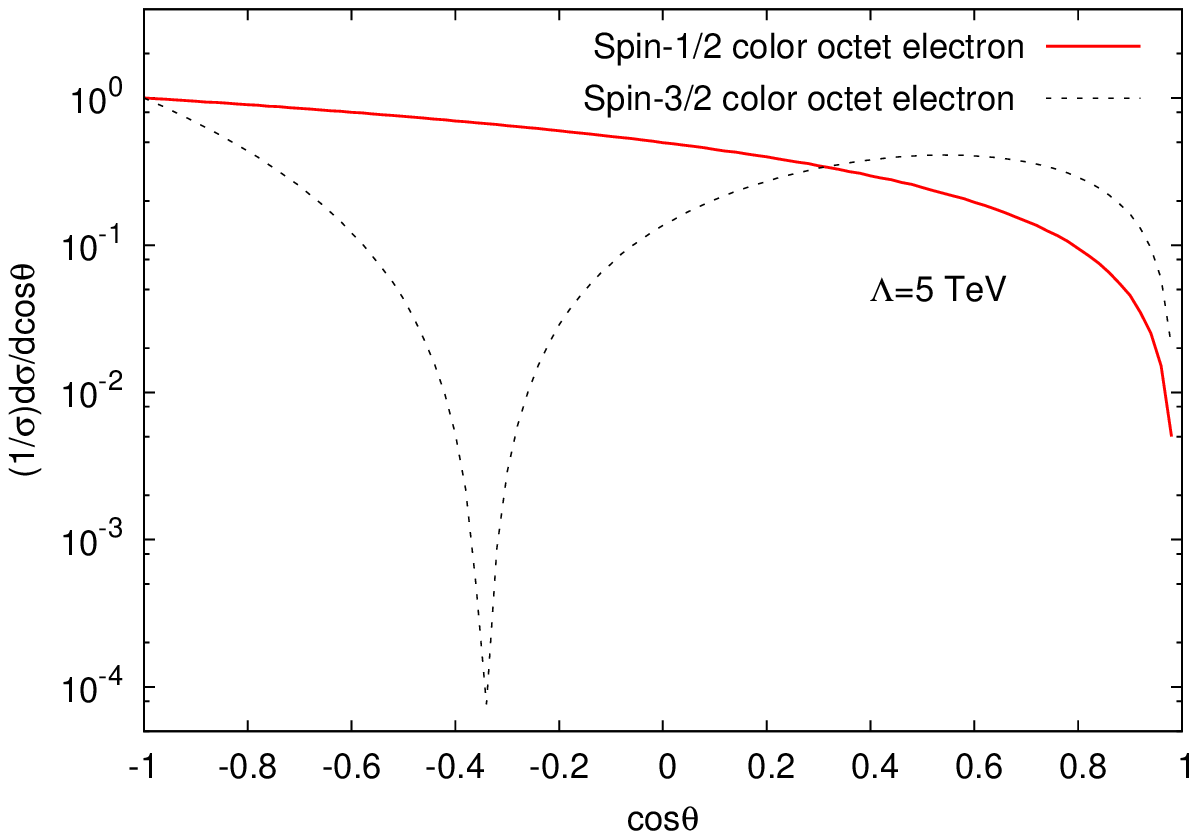}

\caption{Left:The differential cross section as a function the scattering angle
for the spin-3/2 color octet electron ($\Lambda=10$ TeV), and the
spin-1/2 color octet electron ($\Lambda=10$ TeV) at the LHeC-2 with
$\sqrt{s}=2.049$ TeV. Right: The differential cross section as a
function the scattering angle for the spin-3/2 color octet electron
($\Lambda=5$ TeV), and the spin-1/2 color octet electron ($\Lambda=5$
TeV) at the LHeC-2 with $\sqrt{s}=2.049$ TeV. In the both panels,
spin-1/2 and spin-3/2 color octet electron mass values are taken as
$1$ TeV. }
\end{figure}

\begin{figure}
\includegraphics[scale=0.7]{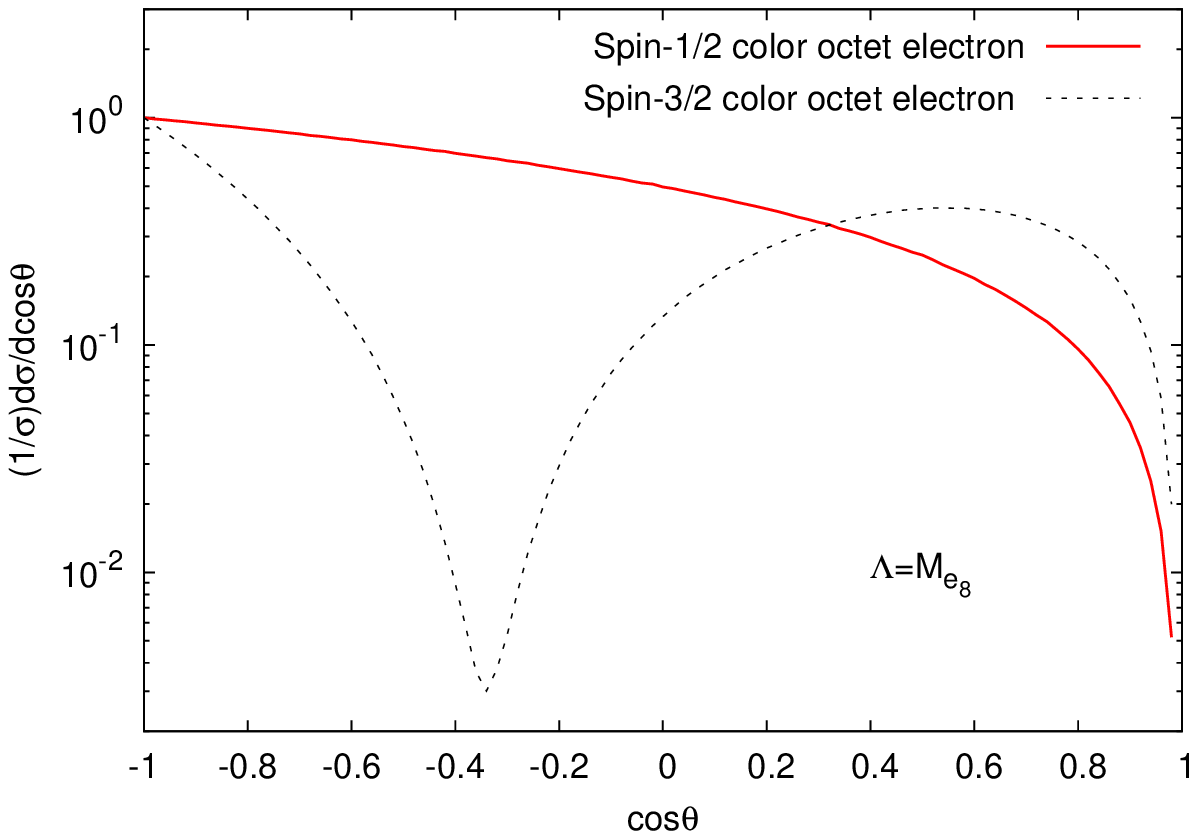}

\caption{The differential cross section as a function the scattering angle
for the spin-3/2 color octet electron ($\Lambda=M_{e_{8}}$), and
the spin-1/2 color octet electron ($\Lambda=M_{e_{8}}$) at the LHeC-2
with $\sqrt{s}=2.049$ TeV. The spin-1/2 and the spin-3/2 color octet
electron mass values are taken as $1$ TeV in our calculations. }
\end{figure}

In order to estimate compositeness scale for spin-3/2 color octet
electron at the LHeC-2 with $\sqrt{s}=2.049$ and $L_{int}=1000\, fb^{-1}$
, we have plot the compositeness scale as a function of spin-3/2 color
octet electron mass in Figure 21. We present reachable values of the
compositeness scale for some color octet electron mass values in Table
5. It can be seen from Table 5, the spin-3/2 color octet electron
with $M_{e_{8}}=1$ TeV can be discovered up to $13.1$ TeV compositeness
scale value at the LHeC-2 with $\sqrt{s}=2.049$ TeV and $L_{int}=1000\, fb^{-1}$.
\begin{figure}
\includegraphics[scale=0.7]{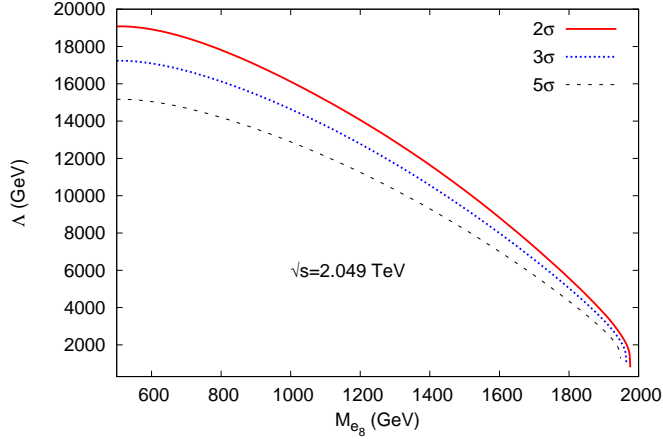}

\caption{Reachable values of the compositeness scale as a function of spin-3/2
color octet electron mass for the LHeC-2 with $\sqrt{s}=2.049$ TeV
and $L_{int}=1000\, fb^{-1}$.}
\end{figure}

\begin{table}
\begin{tabular}{|c|c|c|c|}
\hline 
\multirow{2}{*}{$M_{e_{8}}$, GeV} & \multicolumn{3}{c|}{$\Lambda$, TeV}\tabularnewline
\cline{2-4} 
 & $5\sigma$ & $3\sigma$ & $2\sigma$\tabularnewline
\hline 
\hline 
$500$ & $15.6$ & $17.7$ & $19.6$\tabularnewline
\hline 
$600$ & $15.4$ & $17.5$ & $19.3$\tabularnewline
\hline 
$700$ & $14.9$ & $17.0$ & $18.8$\tabularnewline
\hline 
$800$ & $14.4$ & $16.4$ & $18.1$\tabularnewline
\hline 
$900$ & $13.8$ & $15.7$ & $17.3$\tabularnewline
\hline 
$1000$ & $13.1$ & $14.9$ & $16.5$\tabularnewline
\hline 
$1100$ & $12.3$ & $14.0$ & $14.9$\tabularnewline
\hline 
$1200$ & $11.5$ & $13.1$ & $14.5$\tabularnewline
\hline 
$1300$ & $10.6$ & $12.0$ & $13.3$\tabularnewline
\hline 
$1400$ & $9.6$ & $10.9$ & $12.0$\tabularnewline
\hline 
$1500$ & $8.5$ & $9.6$ & $10.7$\tabularnewline
\hline 
$1600$ & $7.3$ & $8.3$ & $9.2$\tabularnewline
\hline 
$1700$ & $6.0$ & $6.9$ & $7.6$\tabularnewline
\hline 
$1800$ & $4.6$ & $5.3$ & $5.9$\tabularnewline
\hline 
$1900$ & $3.0$ & $3.5$ & $3.9$\tabularnewline
\hline 
\end{tabular}

\caption{Reachable values of the compositeness scale for some spin-3/2 electrons
mass values at the LHeC-2 with $\sqrt{s}=2.049$ TeV and $L_{int}=1000\, fb^{-1}$.}
\end{table}

\section{CONCLUSION}

We have performed a search for resonant production of spin-3/2 color
octet electron at the LHeC. We show that if the compositeness scale
($\Lambda$) equals to $5$ TeV, LHeC-1 with $\sqrt{s}=1.296$ and
$L_{int}=10\, fb^{-1}$ will give the opportunity to discovery chance
up to $M_{e_{8}}=660$ GeV and observation chance up to $M_{e_{8}}=777$
GeV for spin-3/2 color octet electron. In addition, LHeC-1 give the
opportunity for exclusion of spin-3/2 color octet electron up to $849$
GeV. When we consider $\Lambda=M_{e_{8}}$ and $L_{int}=10\, fb^{-1}$case,
upper mass values of the spin-3/2 color octet electron are $1.19$
TeV for discovery ($5\sigma$), $1.21$ TeV for observation ($3\sigma$)
and $1.22$ TeV for exclusion ($2\sigma$).

LHeC-2 with $\sqrt{s}=2.049$ TeV and $L_{int}=1000\, fb^{-1}$ will
give the opportunity to discovery ($5\sigma$) up to $M_{e_{8}}=1.36$
TeV, observation ($3\sigma$) up to $M_{e_{8}}=1.47$ TeV and exclusion
($2\sigma$) up to $M_{e_{8}}=1.55$ TeV for spin-3/2 color octet
electron with $\Lambda=10$ TeV. At the LHeC-2 with $\sqrt{s}=2.049$
TeV and $L_{int}=1000\, fb^{-1}$, these upper mass values are $1.77$
TeV for discovery ($5\sigma$), $1.82$ TeV for observation ($3\sigma$)
and $1.85$ TeV for exclusion ($2\sigma$) of spin-3/2 color octet
electron with $\Lambda=5$ TeV. For spin-3/2 color octet electron
with $\Lambda=M_{e_{8}}$, we find out that the discovery limit ($5\sigma$)
is $1.95$ TeV, the observation limit is $1.97$ TeV and the exclusion
limit is $1.98$ TeV at the LHeC-2 with $\sqrt{s}=2.049$ TeV and
$L_{int}=1000\, fb^{-1}$.

We also show that LHeC-1 with $\sqrt{s}=1.296$ TeV and $L_{int}=10\, fb^{-1}$
will give the opportunity for discovery ($5\sigma$) of $e_{8}$ with
$M_{e_{8}}=500$ GeV up to $\Lambda=5.68$ TeV. We will observe ($3\sigma$)
$e_{8}$ with $M_{e_{8}}=500$ GeV up to $\Lambda=6.47$ TeV and exclude
($2\sigma$ ) $e_{8}$ with $M_{e_{8}}=500$ GeV up to $\Lambda=7.17$
TeV at the LHeC-1 with $\sqrt{s}=1.296$ TeV and $L_{int}=10\, fb^{-1}$
(see Table 3). LHeC-2 with $\sqrt{s}=2.049$ TeV and $L_{int}=1000\, fb^{-1}$
will give opportunity for discovery of $e_{8}$ with $M_{e_{8}}=500$
GeV up to $\Lambda=15.6$ TeV. We will observe $e_{8}$ with $M_{e_{8}}=500$
GeV up to $\Lambda=17.8$ TeV and exclude $e_{8}$ with $M_{e_{8}}=500$
GeV up to $\Lambda=19.6$ TeV at the LHeC-2 with $\sqrt{s}=2.049$
and $L_{int}=1000\, fb^{-1}$. LHeC-2 with $\sqrt{s}=2.049$ and $L_{int}=1000\, fb^{-1}$
will give opportunity for discovery ($5\sigma$) of $e_{8}$ with
$M_{e_{8}}=1$ TeV up to $\Lambda=13.1$ TeV. We will observe $e_{8}$
with $M_{e_{8}}=1$ TeV up to $\Lambda=14.9$ TeV and exclude $e_{8}$
with $M_{e_{8}}=1$ TeV up to $\Lambda=16.5$ TeV at the LHeC-2 (see
Table V). 
\begin{acknowledgments}
I would like to thank Saleh Sultansoy and Gökhan Ünel for their useful
discussions and helpful comments. This work is supported by TUBITAK
BIDEB-2219 grant.\end{acknowledgments}

\end{document}